\documentclass[a4paper,12pt]{amsart}

\usepackage[utf8]{inputenc}
\usepackage{graphics}
\usepackage[english]{babel}
\usepackage{natbib}
\usepackage{url}

\usepackage{pdflscape}

\usepackage{acronym}
\usepackage{colortbl}
\usepackage{longtable}

\usepackage{parskip}

\def \secAttrTitle {\textit{Attribute}}
\def \secToolTitle {\textit{Tool}}

\def \ConceptSHa {Scope}
\def \ConceptSHb {Difficulties faced by VI Users}
\def \ConceptSHc {Existing Products and ongoing Research}
\def \ConceptSHd {Necessary Future Research and Consideration}

\def \pwnCellWidth {1.5cm} 
\def \nonPWNcellWidth {11cm} 
\def \refNumCell {2cm}

\begin{document}
	
\sloppy


\title[Literature Review of Computer Tools]{Literature Review of Computer Tools for the Visually Impaired: a focus on Search Engines}


\author{Guy Meyer}
\address{Guy Meyer - M.A.Sc Software Engineering, Computing and Software, McMaster University}
\email{meyerg@mcmaster.ca}
\urladdr{https://guymeyer.me/}

\author{Alan Wassyng}
\address{Dr. Alan Wassyng - Associate Professor, Computing and Software, McMaster University}
\urladdr{https://www.eng.mcmaster.ca/people/faculty/alan-wassyng}

\author{Mark Lawford}
\address{Dr. Mark Lawford - Chair, Computing and Software, McMaster University}
\urladdr{https://www.eng.mcmaster.ca/people/faculty/mark-lawford}

\author{Kourosh Sabri}
\address{Dr. Kourosh Sabri - Founder and Director, McMaster Paediatric Eye Research Group (McPERG), McMaster University}
\urladdr{https://surgery.mcmaster.ca/bio/kourosh-sabri}

\author{Shahram Shirani}
\address{Dr. Shahram Shirani - Professor, L.R. Wilson/Bell Canada Chair in Data Communications, McMaster University}
\urladdr{https://www.eng.mcmaster.ca/ece/people/faculty/shahram-shirani}

%
%
%
%

\renewcommand{\shortauthors}{Meyer, et al.}


\begin{abstract}

A sudden reliance on the internet has resulted in the global standardization of specific software and interfaces tailored for the average user. Whether it be web apps or dedicated software, the methods of interaction are seemingly similar. But when the computer tool is presented with unique users, specifically with a disability, the quality of interaction degrades, sometimes to a point of complete uselessness. This roots from one's focus on the average user rather than the development of a  platform for all (a golden standard). This paper reviews published works and products that deal with providing accessibility to visually impaired online users. Due to the variety of tools that are available to computer users, the paper focuses on search engines as a primary tool for browsing the web.  By analyzing the attributes discussed below, the reader is equipped with a set of references for existing applications, along with practical insight and recommendations for accessible design. Finally, the necessary considerations for future developments and summaries of important focal points are highlighted.
\end{abstract}



\maketitle


\tableofcontents


\section{Introduction}\label{intro}


This literature review encompasses a collection of attributes of common tools for visually impaired internet users. While specific attributes address the major components of computer interfaces, they are not limited to the visually impaired and can be further extended to visually enabled users as well. The attributes, as seen in later sections (Section \ref{primary_attr}), are presented as disjoint topics, allowing the reader to identify key concepts of interest and focus on collected insights. Under each attribute heading there is a total of four subsections that help the reader understand the major issues and progress in the field. \\

With growing demand for internet applications, there is a growing need for integrating visually impaired users to the web. In fact, both sighted and non-sighted users would benefit from more accessible and adaptable interfaces. \\

Many ongoing efforts are actively attempting to bridge the gap between sighted and non-sighted online users, but there still remains a lack of standardization for either end-user. Though users have the flexibility to operate on whichever platform they desire (whether a specific Operating System or Web Browser), no application or device is found to be an obvious choice for the disabled. \\

In addition to the lack of tooling, very few relevant literature reviews was identified addressing issues for online computer tools. Though most research groups are invested in developing technology, few resources are allocated to congregating information of existing efforts in the field. This review attempts to do exactly that; collect information about an array of existing products, projects, and research to present it in a quick access format. \\

While the concepts are presented in written composition, the reader has the option of accessing the same information in a tabular format, found in the supplementary material. The designed purpose of this review is to provide the reader with sufficient background knowledge so confidently initiate their research and developmental efforts. 

\section{Research Methods} \label{research_methods}

	With a wide range of existing research in the field of visual impairment, the scope was limited to reviewing web and computer based assistive devices. The initial realignment of scope was due to the abundance of research efforts, along with a lack of existing systematic reviews. A collection of recognized databases were identified that could serve as reliable sources of peer-reviewed papers. \\

These databases can be categorized into three sectors; Health Science, Engineering, and Other. The Health Science databases are medically focused and contain disability related studies along with diagnostics. Health Science databases include \textit{OVID Medline}, \textit{PubMED}, \textit{Embase}, and \textit{PsycINFO}. The Engineering related databases, focusing more on products, devices, and applications are \textit{Engineering Village}, comprised of \textit{Compendex} and \textit{Inspec}. For additional content, labelled as Other, \textit{Google Scholar} is the desired tool. \textit{Google Scholar} would often feature papers based on popularity, and has an intuitive interface, the user can often encounter inconsistency with regards to the repeatability of a search. \\

Once relevant databases are established, appropriate keywords would help further locate studies of interest. Several keywords were identified to be of significant use since they narrow the search substantially; \textit{Visually Impaired}, \textit{Visual Impairment}, \textit{Blindness}, \textit{Internet}, \textit{web-based}, \textit{Human-Computer Interfaces}, \textit{Search Engine}, \textit{optimization}, and \textit{virtual}. Note that the use of the keyword \textit{`Blind'} generated poor search results due to its association with Blind Studies and Double-Blind Experiments. \\

Other strategies for finding quality publications are by reviewing the reference list of different papers. By browsing the references of each paper, the reader can discover cited works that may advance their own research. Additionally, the reader will often encounter works that are repeatedly cited in separate papers, indicating importance and relevance. \\ 

In this review some studies with significant relevance were analyzed throughout, while others were judged on the information presented in the abstracts and conclusions. Papers deemed as irrelevant were rejected based on three criteria metrics; the year of publication, content relevancy, and release of newer studies (where certain research groups would produce newer comprehensive results).

\section{Tools} \label{tools}

\subsection{What are Tools?}

	A tool can be described as any computer program or application available to the user. This definition is an extension of the variation used by \cite{powsner_navigating_1994}, ``nav tools", concerned with web accessbility. For the purposes of this review the term navigational tools is excellent since Search Engines do exactly that. Furthermore, the idea of client-server software is mentioned, relating well to the nature of most websites as an attempt to offload from the client.\\

Powsner and Roderer also iterate that ``The Internet is not `arranged' in the usual sense of the word" \cite{powsner_navigating_1994}, implying that through its development over decades certain demographics are disregarded (an important point-of-interest in HCI). \\

Some examples of tools are, 

\begin{itemize}
	\item Search Engines
	\item Web Browsers
	\item Word Processors 
	\item Social Media Platforms
	\item Music Applications
	\item Other Human-Computer Interfaces (via mobile or PC)
\end{itemize}

The attributes that will be described in following sections (Section \ref{primary_attr}) are the fundamental components of most computer tools and apps. By understanding how each attribute influences the user, developers can create more intuitive and robust tools.

\subsection{\secToolTitle: Search Engines (SEs)}
\label{section_search_engines}

This review primarily focuses on the application of web-based Search Engines as a tool for computer users with visual impairment. This particular focus is largely due to the high daily dependance on SEs by computer users. Furthermore, there is benefit and independance when when learning to use SEs efficiently. \\
 
The ability to locate desired information online is very useful. But since the internet is so large and complex, the user employees a \ac{SE} to sift through potential results and rank them in relevancy. SEs provide a quick and accurate response to most general knowledge questions along with help in online navigation. \\
	
The concept of SEs is to provide the user with a `glimpse' of a web page, along with bits of relevant information. By analyzing this response the user should be knowledgeable enough on the `potential' of the web page (Webpage Potential - describes how likely is it that this website will be useful in answering the user's query) in order to decide if it is worth delving deeper. \\

The issue for VI users is the inability to quickly and accurately capture the glimpse of the webpage. Additionally, a standard SE relies heavily on the input query in order to retrieve relevant results. Through the understanding related search terms and proper Boolean Logic (such as AND, OR, and NOT) the SE will provide links that are more accurate to answering the initial query \cite{yang_improved_2012}. These additions to standard search methods allow the user to narrow the search space, and as a result, focusing their efforts and reducing the amount of time they spend exploring results \cite{ tsai_mobile_2010}.  \\
	
\ac{SE} are extremely useful tools since they help users congregate a collection of relevant sites and data, otherwise difficult to locate. \\

As for current SE demographics, Google, Baidu, and Bing handle 74.80\%, 11.32\%, and 8.08\% of the world's search queries, repectively \cite{netmarketshare_search_2019}. Furthermore, Google handles approximately 80.79\% of searches on mobile devices, along with 85.43\% of searches submitted by tablets \cite{netmarketshare_search_2019}. These dominant statistics emphasize the need for a small set of assistive tools to aid the visually impaired. \\

In order to help VI users, various applications have been released to quicken the search process. \citeauthor{yang_improved_2012}, have created a \ac{SSEB} that breaks down the \ac{SERP} \cite{yang_improved_2012}. The paper also references an application by Google called Personalized Search which returns more relevant results to the \ac{SERP} by basing current searches on past ones performed by the user \cite{yang_improved_2012}. By employing powerful \ac{API} provided by the major SEs (ie. Custom Search Engine by Google or the Bing Custom Search by Microsoft) the developers do not need to reimplement these alogrithms. Google's PageRank, RankBrain, and Hummingbird search engine algorithms are intricate search techniques that would require lots of effort to recreate. As a result, when developing new tools it is recommended that the focus remains on elevating the user experience rather than optimizing the search results. \\

A different application called \ac{WoW} changes the SERP by tailoring it specifically to the user \cite{ mele_beyond_2010}. VoiceApp is a speech-based web search engine developed by \citeauthor{ griol_voiceapp_2011}. Another useful application is TrailNote that manages the search process for each user to support ``complex information seeking" \cite{ sahib_evaluating_2015}. The use of trail-managers is strongly recommended. Once implemented properly, VI users can focus on synthesizing the information at hand rather than memorizing past results. It is also important for the users to have quick accessibility to their trail, regardless of their proficiency level. \\

In addition, several studies have been published research regarding the effectiveness of SEs, along with helpful concepts. \citeauthor{ tsai_mobile_2010}, focuses on query specification and the minimization of the search space to improving the quality of the SERP \cite{ tsai_mobile_2010}. The paper also identifies the differences between novice and expert searchers \cite{ tsai_mobile_2010}. Other papers study the level of brain activity while using an SE \cite{ small_your_2009}, preferred engines amongst users \cite{ hu_investigating_2015}, and principal components (\ac{PCA}) that construct a standard SE search \cite{ ivory_search_2004}\cite{ tsai_mobile_2010}. Several studies published results on SE metrics \cite{ aliyu_google_2014}, ideal design \cite{ andronico_improving_2006}\cite{ baguma_web_2008}, accessibility evaluation \cite{ lewandowski_accessibility_2012}, and conformance levels \cite{ andronico_can_2004}. Finally, a study by \citeauthor{ sahib_accessible_2012}, has been published documenting how VI users navigate an SE and how they collect information online \cite{ sahib_accessible_2012}. \\
	
\ac{SE} are applications that will only increase in their commonality due to their ability to reduce the work load of the user. It should be clear that VI users would benefit substantially from highly accessible SEs. Furthermore, sighted users would benefit equally with more efficient SEs. As a result, future developments and research should focus on conformance and adequate design to ensure global accessibility to all user types. Developers should also leverage user-oriented techniques since a user's context, history, or trail can impact future searches

\section{Primary Attributes of Tools} \label{primary_attr}

A collection of attributes is described below that are meant to highlight the major elements of computer use for all user types. By accounting for an array of components in the user experience, the reader is able to focus on the concepts that are relevant to their research. \\

The order in which the attributes are presented is effectively arbitrary, meaning that the reader can analyse attributes that are relevant to their application. The reader also has the option of accessing the information organized by reference which can be found in the supplementary material. The appendicies also include direct quotes and generic summaries of specific papers.

\subsection{Section Outline}

Each attribute is discussed under the following subheadings:
\begin{itemize}
	\item \ConceptSHa
	\item \ConceptSHb
	\item \ConceptSHc
	\item \ConceptSHd
\end{itemize}

Note that these subheadings aid in separating the concerns of the reader to allow for additional organization. The reader will also find analysis of important papers embedded in the text to explain the effectiveness of the attribute.

\subsection{\secAttrTitle: Navigation} \label{section_navigation}

\begin{enumerate}
\item \textbf{\ConceptSHa}
The internet being a fantastic source of information it is primarily useful for those who know what they are specifically browsing for and feel comfortable with the information. If a user is proficient in their ability to navigate between webpages then it would be natural for that user to skilfully locate important information. The ability to navigate through a computer system or the internet is an invaluable skill that is being taught at increasingly younger ages. Even more so, computer proficiency is a common requirement when applying for most jobs. 

\item \textbf{\ConceptSHb}
Since most computers, along with their peripherals (mouse, monitors, keyboard, etc.) are designed for sighted users, universal accessibility is not highly prioritized. As a result, the internet is less accessible for non-sighted users that rely on these devices for navigation(more detail regarding information accessibility in Section \ref{section_info_access}). In addition, most webpages are designed to be used as \ac{GUI}s which heavily favour visual elegance over simplicity in navigation. Consequentially, it is very difficult for non-proficient visually impaired internet users to interact with the web, resulting in a less stimulating, slower online experience. \\

A common difficulty found by many visually impaired users is virtual disorientation \cite{ baguma_web_2008}\cite{ismail_search_2010}. This may results from several situations:

\begin{enumerate}
	\item Inability to recall current virtual location\footnote[2]{\textbf{Analogy:} Imagine the internet to be a physical interface that a person can traverse (similar to how a sighted user analyzes a single webpage on a screen). This is analogous to a person walking across their home. Since the space (their home) has been thoroughly navigated by the person it would be trivial for them to complete their journey. But if the space (or interface) is foreign to the person (ex. stranger's home) then the lack of sight would result in a significant disadvantage. VI users, both in the physical and cyber worlds, have to remember where they have been (ie. their Trail) \cite{ sahib_investigating_2014}, their current \textit{location}, and where they intend to go. As a result, VI users would benefit from assistive devices that manage their Trail and aid in the conceptualization of the internet.}: \textit{The website currently observed by user \underline{or} the user's location within a web page.}
	\item Inability to recall previously visited webpages (Known as ``The Trail" \cite{ sahib_evaluating_2015}): \textit{The recent web pages previously visited by user that are relevant for the current session online.}
	\item Indecisiveness regarding future steps: \textit{The websites that the user should visit next.}\\
\end{enumerate} 

Inexperienced users generally cope with this issue by refreshing the web page, restating the search or closing the browser in order to restart (more in section \ref{section_emotional_implications}) \cite{ murphy_empirical_2008}. This dramatic course of action commonly discourages and frustrates the user, since they are forced to retrace their virtual trail.
 
\item \textbf{\ConceptSHc}
The issues with online navigation, as mentioned above, are important to consider due to their strong impact on the user's experience. Technologies have been developed to resolve some of these problems. \citeauthor{yang_improved_2012}, developed a \ac{SSEB}, made to assist with user orientation and access for those who struggle online \cite{yang_improved_2012}. The paper also provides guidelines when adding shortcuts to an application. \citeauthor{ hakobyan_mobile_2013}, have developed the AudioBrowser, used to navigate the web on the go \cite{hakobyan_mobile_2013}. \ac{WoW} developed by \cite{ mele_beyond_2010} creates a single browsable page that can be easily accessed by VI users \cite{ mele_beyond_2010}. \\

The most commonly used application for navigation is JAWS, developed by Freedom Scientific \cite{jaws_overview_2019} \cite{jaws_overview_2019}. This application is compatible with most Windows applications, including web browsers. Users can browse the web using their preferred web browser and special JAWS key commands \cite{jaws_keys_2019}. The user receives web information through dictation synthesized by the JAWS application. \\

If using a Mac computer running the macOS by Apple, a helpful tool for web navigation is called the Rotor that comes installed with \ac{VO} (the native Apple screen reader) \cite{apple_rotor_2017}. This tool attempts to summarize the links, headings and other page elements into groups. So instead of `tabbing' around from link to link, the rotor presents all common links in a single menu to increase navigational ease. These different groups are then presented in adjacent menus in the Rotor. This feature can be used on individual sites as well, congregating information into groups to assist the user. This feature is excellent for mac users with visual impairment that want more from VoiceOver. \\

The VoiceApp \cite{ griol_voiceapp_2011} and the Homer Web Browser \cite{ dobrisek_voice-driven_2002} offer navigation using voice commands alone, the results are also returned via audio \cite{ dobrisek_voice-driven_2002}\cite{ griol_voiceapp_2011}.  The Audio Hallway \cite{ schmandt_audio_1998} provides navigation using head motions where the user passes `rooms' as potential selection options. The physical movement allows the user to be immersed in the online experience, resulting in more control and focus. \\

For users that require multi-session tasks (online tasks that cannot be finished in one sitting), applications such as Search Trail and TrailNote \cite{ sahib_investigating_2014} were developed. They are particularly handy when the user wants to pause their current session, save relevant information locally, and be able to pickup where they left off once they resume their online activity. There is high importance for managing users between sessions since users may forget the mental map they worked hard to create in their previous session. This idea is also applicable with the use of relevance feedback \cite{ tsai_mobile_2010}, that allows the user to draw information from previous sessions. \\

 As applications become increasingly complex, users are expected to keep up with the versatility of these tools. Even more so, users are expected to work on computers for more than just searching the web. The idea introduced by \citeauthor{ sahib_investigating_2014} that addresses `the Trail' \cite{ sahib_investigating_2014} is a powerful concept that highlights the difficulty of VI users to perform long-term computer tasks. This idea stretches to all types of tools, and as a result allows the user to focus on other components of the task at hand. \\

A portion of active research is dedicated to collecting feedback from the user on desirable features. Common requests are the addition of more feedback from the web application to the user \cite{ murphy_empirical_2008}, along with an overview and general hints as to where the user is located virtually on the page \cite{ baguma_web_2008}\cite{ murphy_empirical_2008}. Additional papers study how users with cognitive disabilities navigate the web \cite{ hu_investigating_2015}, how VI users collect information online \cite{ sahib_accessible_2012}, and what elements are leveraged by VI users to aid in their navigational processes \cite{ivory_search_2004}. 

\item \textbf{\ConceptSHd}
There is a strong need for standardization! Even though several navigation software circulating in the market, they tend to develop their own set of commands and shortcuts that the user must memorize to become proficient. Even more so, new developments should aim to minimize the number of commands so that the user is not overwhelmed. Also, the user could simply begin their online tasks in an efficient and intuitive manner. VI users could benefit from an application that would provide a general overview of a web page and allow the user to skim the page similar to sighted users.\\ 

The concept of navigation is closely related to Search Engines (section \ref{section_search_engines}) and Latency (section \ref{section_latency}).
\end{enumerate}

\subsection{\secAttrTitle: User Interface} \label{section_UI}

\begin{enumerate}
\item \textbf{\ConceptSHa}
When designing a product that aids in overcoming a disability it is crucial that the technology prioritizes the user. Too often are devices designed and tested by visually capable developers that seem to be counter-intuitive for VI users, in practical settings. The purpose of assistive technology is to allow full accessibility to those in need without compromising the quality of information and the ease of accessibility.

\item \textbf{\ConceptSHb}
Technologies that can be categorized with poor user interfaces are most noticeably those that neglect a crucial phase in the user's life-cycle: early learning stages. If a technology is complex in nature, then the average user is less likely to rely on its recurring usage.\\

When considering usage of search engines (more in section \ref{section_search_engines}) by visually capable users, the level of simplicity often goes unnoticed. It is the responsibility of the designer to create an interface that conveys the same level of desired intuition as its graphical duality. This boils down to the ability of the designer to implement their interface in a format that could either be used by all or has the capability of transforming into a simplier interface.\\

The JAWS screen reader is an extremely common computer program useful in all computer tasks that an \ac{OS} could offer. A noticeable drawback from its design is the amount of keyboard commands that are available to the user \cite{jaws_keys_2019}. This results in a steep learning curve that must be overcome to achieve proper user proficiency \cite{andronico_improving_2006}\cite{ murphy_empirical_2008}. Additionally, the user must memorize commands which map a keystroke to a `physical' change on the screen (ie. buttons for scrolling or jumping between menus or text blocks). This issue forces the user to draw implicit assumptions regarding explicit changes on the screen, showcasing a poor conversion between visual and non-visual interfaces for the same application.\\ 

There are multiple screen reader options that are built custom for a specific \ac{OS}. \ac{NVDA} is a free screen reader \cite{nvda_2019} that is a part of an initiative to provide access for technology for all. Microsoft narrator is an additional option for those working with the Windows OS \cite{microsoft_narrator_2019}. The tool is turned on with a keystroke combination available at all times. Users may prefer certain screen readers simply due to their key commands or the intonation of synthesized voice.

Visual authentication interfaces also pose difficulties for VI users. A common automated Turing Test service, CAPTCHA, requires a visually capable user to select or decipher components of images in order to prove the user is human. Although these tests are trivial for sighted users, they are nearly impossible if a person struggles with their vision \cite{ murphy_empirical_2008}. 

\item \textbf{\ConceptSHc}
Though user interfaces are related to accessibility, there is a clear distinction between the them (GUI, AUI, TUI) when evaluating the user's operation within an application. The use of \ac{GUI}s is extremely common since it is simplest for sighted users. Unfortunately, the \ac{GUI}s are complex for VI users \cite{ chiang_computer_2005} due to their high visual dependency. As a result several studies have introduces other modalities that could be useful for VI users. The use of \ac{AUI}s, \ac{TUI}s, and combinations of all three (multi-modal systems) are commonly mentioned in the literature \cite{ dobrisek_voice-driven_2002} \cite{frauenberger_mode_2005} \cite{ hakobyan_mobile_2013} \cite{ macias_adaptability_2002} \cite{ schmandt_audio_1998} \cite{ siekierska_internet-based_2008} \cite{ trippas_spoken_2016} \cite{ yang_specialized_2007} \cite{yang_improved_2012}. Other products may not specifically acknowledge the application of a specific \ac{UI}, although their developments generated a unique non-graphical interface. Screen readers are non-GUIs (such as JAWS) that offer accessibility to users on the entire spectrum of \ac{VI}.\\

Several papers study user interfaces and to develop guidelines \cite{ baguma_web_2008} \cite{ chen_detecting_2003} \cite{frauenberger_mode_2005} \cite{ murphy_empirical_2008} \cite{ tsai_mobile_2010} \cite{yang_improved_2012}, statistics \cite{chiang_computer_2005} \cite{ crossland_smartphone_2014} \cite{yang_improved_2012}, or evaluations \cite{halimah_voice_2008} \cite{ macias_webtouch:_2004} \cite{menzi-cetin_evaluation_2017}\cite{ murphy_empirical_2008}\cite{ muwanguzi_coping_2012}\cite{ trippas_spoken_2016} to improve the usability and intuition behind their respective applications. This extends to proper query formulation for search engines \cite{ tsai_mobile_2010} or the acknowledgement of user's level of experience when developing applications (Net Savvy vs. Net Naive) \cite{ small_your_2009}. \\

Many product developments have also been well documented in the literature. JAWS is among the most common UIs for VI users. Unfortunately, its complexity is documented resulting in a high learning curve \cite{ murphy_empirical_2008}. An interesting result noted by \citeauthor{menzi-cetin_evaluation_2017}, is the high preference of JAWS users towards Internet Explorer (IE) web browser \cite{menzi-cetin_evaluation_2017}. The issue with IE is the lack of online community support for the browser. With sighted users the popularity of Internet Explorer is known creating a gap in development and support between VI and sighted users, where sighted users are tailored to and use platforms with long-term support. Other UIs for web accessibility are Mg Sys Visi \cite{halimah_voice_2008}, \ac{KAI} \cite{ macias_adaptability_2002} \cite{ macias_webtouch:_2004}, \ac{WoW} \cite{ mele_beyond_2010}, and the Homer Web Browser \cite{ dobrisek_voice-driven_2002}. Some applications like EasySnap developed by \citeauthor{ jayant_supporting_2011}, aid VI users with developing skills in photography as well as sharing their content online \cite{jayant_supporting_2011}. \citeauthor{siekierska_internet-based_2008}, developed a product to provide users with an interface for physical world navigation, allowing them to use maps freely \cite{ siekierska_internet-based_2008}. \citeauthor{ sahib_evaluating_2015}, have developed a non-visual spelling support system \cite{ sahib_evaluating_2015}. Finally, Audio Hallway is a conceptual \ac{AUI} product developed for browsing collections using head motions, giving the user an immersive experience \cite{ schmandt_audio_1998}.

\item \textbf{\ConceptSHd}
Developers would benefit greatly from referencing and considering these principles when developing applications. The user should not be frustrated with the UI, because if developed with all users in mind these interfaces will become as simple as using a screen. 
\end{enumerate}

\subsection{\secAttrTitle: Information Accessibility} \label{section_info_access}

\begin{enumerate}
\item \textbf{\ConceptSHa}
The process of collecting and synthesizing information from the internet is an important skill to have in order to become efficient in using online applications. But for synthesis to occur, the information must be quickly and easily accessible to the user. Navigation assistance (section \ref{section_navigation}) is not enough to interact with information online, the user must also be able to understand and access the media they encounter. 

\item \textbf{\ConceptSHb}
The internet is designed for sighted users resulting in highly graphical presentation of information. Furthermore, there is little consideration for VI users that may be equipped with screen readers or assistive aids \cite{ macias_webtouch:_2004}. Consequentially, the VI user may read a web page while having to subconsciously guess the contents of information that is inaccessible to them.\\

As an example, consider a university web page terminal that allows students and staff to check for events and updates around campus. Studies have compared a collection of university sites that are ideally supposed to be accessible to all students and yet include surprising levels of inaccessibility \cite{ harper_quest_2008} \cite{menzi-cetin_evaluation_2017} \cite{ muwanguzi_coping_2012}. After evaluating the compliance levels of each site it becomes clear that most visually impaired students cannot access a substantial percentage of university content. This results in lack of knowledge and frustration for the students. In addition these sites did not comply with the \ac{WAI} published by the \ac{W3C} \cite{consortium_web_1999}.\\

It is also important to avoid overloading the user when they are browsing for content \cite{ murphy_empirical_2008}. Since it is quicker to skim through a document visually, it is expected that the online experience is fast. But in the case of VI users the experience may be slowed down to accommodate for screen readers. If the technological aid reads an excessive amount of information from the web page, then the user will experience a slower consumption rate (the rate at which a user is presented with new information). Conversely, if the aid outputs lots of audio, then the user may feel overwhelmed and is forced to slow down equally.

\item \textbf{\ConceptSHc}
The ability to provide accessibility to computer and online apps is increasingly important since virtual media (ie. text, images, videos, etc.) is how the relevant data is commonly represented. As a result, much effort is in converting standard sites to become accessible. As a humanitarian effort to ensure accessibility, the government of Ontario (Canada) has filed the \ac{AODA} detailing standards and regulations that organizations and individuals should abide to make their products or services more accessible \cite{aoda_2019}. The \ac{AODA} also provides good teaching, coding, and design practices that extend past the web to improve the accessibility of public places and schools.  \citeauthor{halimah_voice_2008}, have developed a translator that can convert HTML to multiple mediums, including; voice output, braille, or text. Due to this versatility, the translator can be employed by an array of users including the elderly and other individuals with ranging disabilities \cite{halimah_voice_2008}. \citeauthor{ macias_adaptability_2002}, developed a product named \ac{KAI} that is composed of two modules. The first, a markup language designed for the blind, \ac{BML} \cite{ macias_adaptability_2002} \cite{ macias_webtouch:_2004}. Their other is an app called WebTouch, a multi-modal web browser used in conjunction with \ac{BML} \cite{ macias_adaptability_2002}\cite{ macias_webtouch:_2004}. \\

Another application, VoiceApp, allows web browsing using voice commands alone \cite{ griol_voiceapp_2011}. The VoiceApp generates markup metadata called VoiceXML that indicates relevant voice information to be transmitted \cite{ griol_voiceapp_2011}. The Web Access Project, developed by \citeauthor{ yang_specialized_2007}, adds captions and audio descriptions to video clips as context for VI users \cite{ yang_specialized_2007}. \ac{SSEB} \cite{yang_improved_2012} also adds to the accessibility of the web, by allowing the user to comfortably search for webpages. Additionally, the paper by \citeauthor{yang_improved_2012}, indicates a minimum requirement claiming that, anyone should be able to understand the contents of any web page \cite{yang_improved_2012}. Though this goal may seem ambitious, it depicts the ideal compliance status of the web. \citeauthor{ chen_detecting_2003}, focuses on web browsing via, ``handheld computers, \ac{PDA} and smart phones" \cite{ chen_detecting_2003}. Their application compartmentalizes web content so that it can be accessed using small form-factor devices \cite{ chen_detecting_2003}. \\

Wearable technology is used as a method of accessing information. AlterEgo, a smart, non-invasive, wearable computer which sits externally around the human vocal cords. AlterEgo allows the user to communicate with computers without audibly voicing a word \cite{ kapur_alterego:_2018}. This provides human-computer interaction that is totally discreet (more in section \ref{section_discreetness}). Other physical products are used to provide VI users with access to physical graphical information, such as maps \cite{ hakobyan_mobile_2013}\cite{ roentgen_impact_2009}\cite{ siekierska_internet-based_2008}. These technologies are not only useful when transporting from one location to another but their use declines once the user becomes familiar with the space, indicating the use of successful learning methods \cite{ roentgen_impact_2009}. \\

Additional studies are focused in researching ways to improve accessibility. Several papers study the issue of overloading the user \cite{ baguma_web_2008} \cite{ murphy_empirical_2008} \cite{ trippas_spoken_2016}. Others indicate that the use of multiple modalities (audio, touch or both) are good ways of replacing graphical information \cite{ chiang_computer_2005} \cite{ griol_voiceapp_2011} \cite{ macias_webtouch:_2004}. \citeauthor{ baguma_web_2008}, have produced a list of requirements that aid developers to assure accessibility \cite{ baguma_web_2008}.

\item \textbf{\ConceptSHd}
in the future, when new web content is generated, it is important to take preemptive measures such as; adding alternative text to images and videos, focusing on web page accessibility, and performing proper testing to ensure accessibility with adequate, non-visual computer peripherals (keyboard only). For products and applications that are developed in this field, it is crucial to remember that users have a range of visual impairments along with other disabilities that could also benefit from their product. Current accessibility applications are complex with a large learning curve that is overwhelming for the elderly or naive online users. Developers must consider what is important in terms of accessibility and what can be omitted. 

\end{enumerate}

\subsection{\secAttrTitle: Latency} \label{section_latency}

\begin{enumerate}
\item \textbf{\ConceptSHa}
The efficiency of the web, more specifically search engines, has allowed internet users to spend significantly less time looking for results. Consequentially, a standard user is expecting quick retrieval. The latency, also known as `search time' \cite{yang_improved_2012}, or task completion time, of a web search is the time difference between initial formulation of query and final intake of information. This could be extended to describe the amount of time a user spends on a website to absorb the information. \\

Note the this attribute is not limited to \ac{SE}s since all computer tools are expected to work quickly. Another reason for Latecy to be regarded as an important attribute is that it attempts to numerically quantify how useful and accessible a tool is. 
	
\item \textbf{\ConceptSHb}
The internet has become an endless pool of knowledge that can be ideally accessed by anyone. The major distinction between VI and sighted users is their ability to consume information quickly. Since a sighted users has higher visual acuity they are comfortable skimming through dense pages with lots of data. Conversely, VI users are forced to examine the same page more carefully, resulting in a slower online experience. Each VI user then spends more time per webpage and therefore experiences more latency between query and result.\\

\citeauthor{ mack_inattentional_1998}, have also addressed the issue of latency but with an attempt to identify its source. They claim that VI users construct explicit perceptions of webpages, rather than visually driven implicit observations \cite{ mack_inattentional_1998}. These explicit perceptions are more difficult to comprehend and force VI users to spend more time online.
	
\item \textbf{\ConceptSHc}
Several studies attempt to quantify the difference in time duration between VI and sighted users. \citeauthor{menzi-cetin_evaluation_2017}, captured the latency of VI users when completing online tasks \cite{menzi-cetin_evaluation_2017}. \citeauthor{ ivory_search_2004}, also focus on speed of information access and collected measurable metrics on user evaluation time \cite{ ivory_search_2004}. Others have highlighted the advantage of using mobile devices \cite{ tsai_mobile_2010}, and the importance of developing more efficient user interfaces for SE users \cite{ andronico_improving_2006}. \\

 Products that have been developed to aid VI users have tackled a variety of issues. AlterEgo, a wearable input device, allows the user to input information to a computer at a faster rate and at any distance since the wearable collects muscle movements directly from the vocal cords \cite{ kapur_alterego:_2018}. EasySnap, an application for sharing pictures and videos, is intuitive and easy to operate \cite{ jayant_supporting_2011}. This allows users to become faster with simple activities, reducing latency in processes such as sharing media. Search Trail, a multi-session assistant for VI users, reduces the resumption time between sessions by allowing the user to revisit their virtual trail and pick up where they left off \cite{ sahib_investigating_2014}. Wikipedia, the online encyclopedia, is a good source for quick descriptions \cite{ griol_voiceapp_2011}.\\

With immediate accessibility to the web, many applications are capable of answering questions, providing guidance and help manage personal devices. Examples of these applications include; ``Siri" by Apple, ``Hey Google" by Google, ``Alexa" by Amazon, and ``Cortana" by Microsoft \cite{yoffie_voice_2018}. Other examples specifically for SEs include the Featured Results at the top of the Google SERP that attempts to determine the most confident result.
	
\item \textbf{\ConceptSHd}
Through the interaction with online apps, users generally favour those that reach solutions quicker. Whether it is the start-up time, resumption time, or time spent completing tasks, developers and engineers must focus on minimizing the latency of the overall experience.
\end{enumerate}

\subsection{\secAttrTitle: Discreetness} \label{section_discreetness}

\begin{enumerate}
\item \textbf{\ConceptSHa}
The usage of computers and assisted devices are often helpful for those that require support due to a disability. But users are most likely to favour devices that allow them to operate in a discreet manner. The user is then free to explore the web as they please, without the fear of stigmatization or the negative social implications of using obtrusive devices. Note that the concept of discreetness extends past the context of computer usage, and into all fields of assistive devices.
	
\item \textbf{\ConceptSHb}
During standard computer usage, in public settings, it often goes unnoticed that a sighted user is capable of using their devices discreetly. The user enjoys privacy via speechless text entries, auditory feedback through headphones, and small sized screens that can be hidden from others. The user can then reduce their noticeability and can blend with the local setting (ie. library, coffee shop, waiting room, etc.). This concept is most prevalent in youth settings, where phone and computer usage is common, while everyone is fixed on their own device. \\ 

A lack of discreetness occurs when others hear and see the user's operation. Since computers are extremely helpful in completing rudimentary tasks, such as; checking emails, sending money, and setting reminders, they may be quite helpful for VI users in their everyday routine. But in order to help the user employ the technology more comfortably it would be beneficial if the operation is physically hidden (or at the least discreet).\\
	
As an example, imagine designing a new controller for VI users. If the device requires the user to swing their arm violently, then it would be rejected by others due to social norms, negatively affecting the individual. \\

So although the device may operate efficiently and accurately, it would still be rejected by the end-user. It is the responsibility of the designer to account for a wide range of usages in order to determine if the device is user friendly in public settings.\\  
	
Finally, for the users to experience adequate UI/UX they should not be required to disclose their handicap. As a result VI users may appear as an ordinary user to others. This form of confidentiality has positive emotional implications (more in section \ref{section_emotional_implications}).
	
\item \textbf{\ConceptSHc}
Discreetness can be expanded into components; user-voicing, and audio feedback. For user-voicing application discreetness is more difficult to achieve since the user is forced to audibly operate the application. For audio feedback applications, the user may use computer peripherals such as a keyboard, mouse or touchpad to silently interact with the computer or device. An example of user voicing applications can be found in \cite{ griol_voiceapp_2011} \cite{halimah_voice_2008} \cite{ismail_search_2010} \cite{ trippas_spoken_2016}. Examples of audio feedback applications can be found in \cite{ dobrisek_voice-driven_2002} \cite{frauenberger_mode_2005} \cite{ griol_voiceapp_2011} \cite{ hakobyan_mobile_2013} \cite{halimah_voice_2008} \cite{ismail_search_2010} \cite{ jayant_supporting_2011} \cite{ kapur_alterego:_2018} \cite{ macias_webtouch:_2004} \cite{ murphy_empirical_2008} \cite{ schmandt_audio_1998} \cite{ siekierska_internet-based_2008} \cite{ trippas_spoken_2016} \cite{ wu_visually_2014}. \\

A wearable, non-invasive device called AlterEgo, is designed to allow the user to communicate with a computer without audibly pronouncing words \cite{ kapur_alterego:_2018}. When using audio feedback devices, VI users must be focused by listening for long periods of time without distraction, the prorgrams also limit a user's ability to scan the page and jump between results (a common use-case in web page search). As a result, the developments of tactile or multi-modal interfaces would be beneficial \cite{frauenberger_mode_2005}. \citeauthor{ siekierska_internet-based_2008}, have developed map interfaces for physical world navigation that focus on tactile interfaces allowing the user to have their ears listening for dangers or physical threats to their commute (ie. cars, other pedestrians, traffic light signals, and more) \cite{ siekierska_internet-based_2008}.  \\


The AlterEgo device developed by \citeauthor{ kapur_alterego:_2018} motivates the possibility of creating a fully discreet system. By allowing the user to quickly send text phrases as input, the system may be operated anywhere in public. Similar to how SMS messages are a totally discreet form of communication, with the help of devices like AlterEgo, VI users will now be able to send and accept computer data without being noticed. The current methods for discreet text entry include a standard `QWERTY' keyboard, mobile keyboards with predictive text, and other forms of non-verbal entries.
	
\item \textbf{\ConceptSHd}
With much dependence on internet accessibility for all indicates that there is a growing need for discreetness in computer/web applications, resulting in quick and quiet access (what is now only present for the average or `ideal' user). Once VI users can comfortably operate their devices and participate in online activities regularly, they will be more likely to develop independence and social awareness.
\end{enumerate}

\subsection{\secAttrTitle: Emotional Implications} \label{section_emotional_implications}

\begin{enumerate}
\item \textbf{\ConceptSHa}
The emotions that a user feels when facing challenges in an unfamiliar environment is important to the success of a product. Whether it is happiness, confusion, frustration, or despair, the response of a user to the functionality of a product is a definite indication of its usability and accuracy.
	
\item \textbf{\ConceptSHb}
Common negative reactions by individuals using web browsers are frustration since the user is unable to determine their virtual location or recall previously acquired knowledge. Depending on the experience of the user, a potential solution would be to close the program and retrace their steps (discussed in section \ref{section_navigation}) \cite{ murphy_empirical_2008}. \\

When content is found to be inaccessible, additional frustrations and confusions can set in. Furthermore, partially accessible web pages are equally problematic since text may be easily understood via screen readers though images and videos are unobservable. This form of inaccessibility is critical since many sites rely on visual content to convey the most relevant info. \\

The use of standard webpages and search engines browsing constitutes for a large part of computer usage. In addition, participating in social media platforms has become a standard for many. Although there are conflicting views on the effects of social media, \citeauthor{ jayant_supporting_2011}, have indicated that social media is beneficial for VI users \cite{ jayant_supporting_2011}. Since social media allows for anonymity in social settings, people with visual impairment can freely express themselves without real-world confrontation \cite{ world_health_organization_who_world_2011}. Additionally, VI users can share their experiences with others almost instantaneously and receive positive feedback from their peers online. 
	
\item \textbf{\ConceptSHc}
Several studies in the literature have focused their research to understand the human factors of accessible products. \citeauthor{menzi-cetin_evaluation_2017}, have noted the importance of usable products since they make its users happy \cite{menzi-cetin_evaluation_2017}, these factors are often overlooked. \citeauthor{ tsai_mobile_2010}, attempts to quantify the amount of time it takes for a VI user to get frustrated while using the web \cite{ tsai_mobile_2010}. \citeauthor{ murphy_empirical_2008}, also indicates the frustration behind online application \cite{ murphy_empirical_2008}, \citeauthor{ andronico_improving_2006}, describes the need for less frustrating user interfaces that are more user-oriented \cite{ andronico_improving_2006}. Other papers will describe the dependency of VI users \cite{ hersen_assertiveness_1995} and the increased brain activity used by Net Savvy (experienced) users \cite{ small_your_2009}, emphasizing the non-intuitive nature of online platforms. \\

The World Health Organization \cite{ world_health_organization_who_world_2011} highlights the importance of online communities as a method to overcome barriers experienced by face-to-face interaction \cite{ world_health_organization_who_world_2011}. \citeauthor{ wu_who_2011}, studied a text-based online social network platform, Twitter, displaying the levels of influence by specific VI groups \cite{ wu_who_2011}. \citeauthor{ wu_visually_2014}, analyzed the social network density, size and usage of VI and sighted Facebook users \cite{ wu_visually_2014}. VI users on Facebook were identified by their use of Apple's iOS voiced accessibility feature, VoiceOver, these users statistically received more feedback from peers \cite{ wu_visually_2014}. \citeauthor{ jayant_supporting_2011}, found that sharing pictures and videos online had a positive effect on the individual \cite{ jayant_supporting_2011}. Other research indicated that VI users feel as though they are missing out on a perceptual experience online \cite{ murphy_empirical_2008}, referencing the stimulus of visual content. In educational setting, \citeauthor{ muwanguzi_coping_2012}, has studied the reactions of VI students when accessing web-based educational content, as well as, their ability to communicate with professors and colleagues virtually \cite{ muwanguzi_coping_2012}. \\

 \citeauthor{ismail_search_2010} present, in 2010, the disappointment of VI users when using voice activated browsers. \citeauthor{ hakobyan_mobile_2013}, studied the motivation behind the development of a \ac{MAT}, discovering that, ``individuals feel less stigmatized or labelled", when using these products. This topic relates well to the concept of discreetness discussed in section \ref{section_discreetness}. Search Trail, an application that aids VI users with multi-session tasks, provides the users with confidence knowing that the program tracks their virtual trail \cite{ sahib_investigating_2014}. The most frequent reasons for not using applications or devices is the lack of interest, cost, or simply being unaware of its availability \cite{ crossland_smartphone_2014}. 
	
\item \textbf{\ConceptSHd}
There is a need for humane considerations of emotional implications while developing future products, devices and applications. It is not enough to make a product that it is accessible, but also one that allows the user to enjoy the online experience. It should not be a burden for the user to interface with online applications but rather an integral part of a person's life. If an application causes users to become frustrated, the likelihood of repetitive use declines dramatically, resulting in abandoned devices and products. Developers and engineers must consider the end user in their entirety. 
\end{enumerate}

\section{Understanding the User Space} \label{section_user_space}

When designing a \ac{UI} theres a constant battle between user freedom and providing an overwhelming amount of information. By binding user commands to keystroke combination the user spends less time while achieving more functionality. But first they must spend time memorizing commands \cite{ murphy_empirical_2008}. This concept of user preference for `recognition over recall' is exemplified by Scott MacKenzie \cite{mackenzie_hci_2012}, in his analysis of menus. So where is the balance? The easy answer is, it depends. Primarily on the users of the application along with their preferences and capabilities. \\

First lets define what the user wants to reach, perhaps everything. Since this review is centralized around search engines lets focus on the Internet. An environment that encompasses all that the web has to offer is known as the \textit{cyberspace}, coined by William Gibson in 1982 \cite{gibson_burning_2014}. This idea encapsulates the struggle of UI design. Since users want to traverse the cyberspace as quickly as possible without memorizing steps.\\

Let's hone it down by concentrating on a single webpage. By forgetting (for a moment) all the places this page could take you, it becomes easier to see how a single webpage is a \textit{defined} cyberspace. More specifically, a webpage has a better defined set of possibilities, being a subset of the total cyberspace. It's like counting the leaves of a tree versus the entire forest.\\

As an example consider the SERP. Being a results page it holds a finite number of search results. By manipulating the way in which the user interacts, they might find a more efficient method of accessing the same information. In the case of VI users, this is definitely true. Primarily due to a large majority of webpages being designed for sighted user. \\

To conclude it is important to consider the users' space of possibilities when designing tools. In the context of SEs the user's end result is unpredictable since the space is infinite. But for other tools like word processors or music applications that allow the user to interact with different content the space may be very well defined. As a result, by looking at the whole picture designers may find interfaces that benefit users of all type regardless of their visual acuity.

\section{Evaluations and Surveys} \label{section_evals}

\subsection{Evaluations}

As a tool developer it is important to consider the usability of your implementation. Simply satisfying the requirements does not necessarily prove success. Structuring evaluations that provide statistical backing to your application can be achieved by focusing on one of the attributes analysed above (section \ref{primary_attr}).\\

For example, examiners can focus on latency as their main quantitative factor. By doing so they are able to select an independent variable more easily. For latency this could be a timed variable. While for information accessibility the metric could be a percentage of inaccessible items in a site. Scott MacKenzie \cite{mackenzie_hci_2012} describes the field of human-computer interaction accurately and provides techniques for structuring, defining, and carrying out evaluations and experiments.

\subsection{Visual Question Answering surveys} \label{section_vqa}

A VQA survey provides its participants with an image. They are then asked specific questions regarding that image so it can be alternatively described. This data is recorded and can later be used to develop vision algorithms that can annotate web content automatically. This technology can provide VI users with a `personal assistant' that answers questions like a human.\\

When considering key webpage components, images and videos are often the entire focus (think of Youtube, Instagram or Facebook).  This can be seen by the UI features like Youtube's Autoplay and Facebook's endless scroll. Successful VQA methods are extremely useful since they allow VI users to generate annotations upon request. Several VQA papers are described in this text that attempt to resolve this issue.\\

Several VQA surveys have been conducted historically, this review congregated a sample of relevant surveys \cite{agrawal_vqa:_2017}\cite{gurari_vizwiz_2018}\cite{wang_explicit_2015}. The success of these algorithms is still unclear due to the complexity and variability of the initial problem. Regardless this issue is quite important since the annotation of every picture online is an impossibly exhaustive task. If this aligns with your field of interest it would be well advised to delve deeper into Machine Learning (ML) algorithms concerned with classification.\\

The results of VQA surveys are useful in understanding what users are interested in knowing. By analyzing commonalities between different results, surveys and content types, researchers can synthesize data to determine common concepts of intrigue. Additionally, predictive software, such as machine learning techniques, can be employed by VI users to answer simple questions discreetly. This provides the user with independence, confidence, and excitement.

\section{Related \& Future Work} \label{related_work}

With regards to future expectations from this research team, a product will be developed that aims to aid VI internet users browse a standard \ac{SERP}. All sections of this review will be further leveraged in the development of a computer tool for SEs. Furthermore, the device will be accompanied by an evaluation that will assess the quality of the computer tool when compared with the participants' preferred method of accessibility. Since the research interests at McMaster University are primarily pediatric, visually impaired students of all ages will ideally participate in this evaluation.\\

McMaster also encourages final year engineering students to focus their projects on Physical World Navigation (PWN). This helps build interest in the field of visual impairement while motivating students to work on real-world problems.

\section{Conclusion} \label{conclusion}

Highly accessible and user-friendly computer tools are difficult to design. Furthermore, due to the wide spectrum of visual disabilities, it is impossible to tailor an ideal program for each user. As a result, designers and engineers must employ the characteristics and attributes that are most concerned with the needs of the target audience. \\

Common results uncovered in this review include; the lack of user-oriented accessible design, incompatibility with existing technologies, and the difficulty of becoming proficient as a regular user (the learning curve). \\

When faced with navigation, VI users are forced to construct abstract layouts of the page since most developments are targeted to the visually enabled. Additionally, information is less accessible than originally perceived, most images and videos include little, if any, alternative descriptive text, while standard menus and options are difficult to find, mostly, to enhance visual attractiveness. \\

The online community lacks standardization causing each individual to build their own distinct portfolio of preferred programs. Furthermore, the interaction of VI users with these interfaces, particularly in public settings, is rather obtrusive and loud. The lack of overall discretion attracts unwanted attention to a demographics that simply wants to use common applications. \\

This review attempts to merge the attributes most commonly addressed in the literature, while focusing on Search Engines (SEs) as the assistive tool. SEs have become a natural starting point for any web search due to their simplicity and usefulness. Unfortunately, these entry-level platforms include ingrained barriers that restrict VI users from browsing freely. \\ 

Sections explaining tools, along with common examples of their use can be found in Tools, section \ref{tools}. More analysis of issues and difficulties attributed to the online experience are described in depth starting in section \ref{primary_attr}, Primary Attributes of Tools.


\appendix

\section{List of Acronyms}
\addcontentsline{toc}{chapter}{List of Acronyms}
\begin{acronym}
	\acro{ADP}{Assistive Devices Program}
	\acro{AODA}{Acessibility for Ontarians with Disabilities Act}
	\acro{API}{Application Programming Interface}
	\acro{AUI}{Auditory User Interface}
	\acro{BLV}{Blind or Low Vision}
	\acro{BML}{Blind Markup Language}
	\acro{CSE}{Custom Search Engine}
	\acro{EA}{Educational Assistant}
	\acro{GC}{Google Classroom}
	\acro{GUI}{Graphical User Interface}
	\acro{HCI}{Human-Computer Interaction}
	\acro{HLD}{High Level Design}
	\acro{HTML}{Hypertext Markup Language}
	\acro{IE}{Internet Explorer}
	\acro{KAI}{Accessibility Kit for the Internet}
	\acro{MAT}{Mobile Assistive Technology}
	\acro{MOS}{Mean Opinion Score}
	\acro{NVDA}{NonVisual Desktop Access}
	\acro{OBS}{Open Broadcaster Software}
	\acro{OM}{Orientation and Mobility}
	\acro{OS}{Operating System}
	\acro{PCA}{Principal Component Analysis}
	\acro{PDA}{Personal Digital Assistant}
	\acro{PWN}{Physical World Navigation}
	\acro{RBD}{Refreshable Braille Display}
	\acro{SERP}{Search Engine Results Page}
	\acro{SE}{Search Engine}
	\acro{SSEB}{Specialized Search Engine for the Blind}
	\acro{STT}{Speech-To-Text}
	\acro{TTS}{Text-To-Speech}
	\acro{TUI}{Tactile User Interface}
	\acro{UIs}{User Interfaces}
	\acro{UI}{User Interface}
	\acro{URL}{Universal Resource Locator}
	\acro{VI}{Visual Impairment}
	\acro{VO}{VoiceOver}
	\acro{W3C}{World Wide Web Consortium}
	\acro{WAI}{Web Accessibility Initiative guidelines}
	\acro{WoW}{WhatsOnWeb}
	\acro{YRDSB}{York Region District School Board}
\end{acronym}


\begin{landscape}
	\section{Supplemental Material} \label{appendices}

The appendicies represent a detailed version of the written text above. Each Concept displayed in a Concept Matrix according to references that compose the section. The reader may utilize this categorical representation to focus on specific papers within a concept. At the bottom of each appendix section there is a list of all references found in the section. \\

Note that different readers may benefit from either of the two dualities. It is recommended to have a general research question or project direction when analyzing the text so to not be overwhelmed. \\

This section categorizes the concepts exemplified by each paper \cite{webster_analyzing_2002}. This will help the reader quickly reference articles that address relevant concepts.

\newpage

\subsection{Search Engines} \label{append_search_engines}

Search Engines

\begin{longtable}{|c|p{7.5cm}|p{7.5cm}|}  \hline
	
	\textbf{Ref \#} & \multicolumn{1}{c|}{\textbf{Products}} & \multicolumn{1}{c|}{\textbf{Studies}} \\ \hline
	
	\cite{leporini_designing_2004} & & ``describe the main design issues affecting the user interface of a search engine when a sightless user interacts by means of a screen reader or voice synthesizer." \newline\newline
	``the most important differences between a visual layout and aural perception" \\ \hline
	\cite{yang_improved_2012} &  \underline{Objective}: \newline
	- Introduce the concept of SSEB (Specialized Search Engine for the Blind) (abstract)
	\newline \underline{Target Goals}: \newline
	- ``to make visually disabled people able to keep pace with the changing World Wide Web and to improve the efficiency of their information searches."
	\newline - Figure 3 captures the major elements of the Google SE results page \newline
	
	[16] Jansen and Spink examined characteristics and changes in Web search from nine studies of five Web search engines based in the US and Europe (p2) \newline
	[23] Topi and Lucas examined the effects of the search interface and Boolean logic training on user search performance and satisfaction the assistive search tool had a positive effect on performance satisfaction and confidence \newline
	[13] Google made a beta function version of their search engine open to public use called Personalized Search... users could get the results most relevant to them based on what they have searched for in the past (p3)
	& \\ \hline
	\cite{malik_oriental_2014} & & \underline{SE mentioned}: Google, Yahoo, Bing, AltaVista, MSN Search, and Ice rocket $\rightarrow$ most of these SE dont exist anymore in 2014 \\ \hline
	\cite{ tsai_mobile_2010} & & - Important: ``very few studies consider narrowing down the search space in the query formulation step" (abstract)
	\newline - Introduces the importance of query specifications during the query step (ie. what type of document are you looking for?) (p1)
	\newline - Only a fraction of retrieved documents are relevant (p1)
	\newline - Figure 1: Shows PCA for information access in web SE (p2) \newline
	
	\underline{Two types of searches}:
	\newline \textit{Novice Searches}: ``users do not have some prior knowledge to search specific information" (p2)
	\newline \textit{Expert Searches}: ``users usually have this knowledge to search related information by some precise keyword" (p2) \newline
	
	Search Space must be minimized $\rightarrow$ Avoid the feedback path in PCA is very important
	\\ \hline
	\cite{ aliyu_google_2014} & & Section 2: Short LR includes SE metrics accessibility 
	
	Explains difference between implicit and explicit search \\ \hline
	\cite{ small_your_2009} & & Discusses the level of brain activity required while using search engines \\ \hline
	\cite{ gurari_vizwiz_2018} & This VQA survey is similar to a Search Engine queries since it takes in a query tuple (question + image) and produces a single solution (most relevant answer to question) & \\ \hline
	\cite{ mele_beyond_2010} &  - WoW changes the SERP (search engine report page) so that it can be modified to form the user. (p2) & \\ \hline
	\cite{ griol_voiceapp_2011} & \underline{VoiceApp}: a complete speech-based web search engine & \\ \hline
	\cite{ hu_investigating_2015} & & \underline{Results}: \newline
	``participants overwhelmingly preferred the search engine method to the two browsing conditions" (abstract)
	\\ \hline
	\cite{ baguma_web_2008} & & \underline{Covers common web applications}:
	\newline - search engines 
	\newline - news portals 
	\newline - e-commerce and a tourism portal
	\\ \hline
	\cite{ sahib_evaluating_2015} & “we use our previous findings to inform the design of a search interface [TrailNote] to support visually impaired users for complex information seeking.” & \\ \hline
	\cite{ lewandowski_accessibility_2012} & & \underline{Objective}: \newline
	``describe the aspects to be considered when evaluating web search engines' accessibility for people with disabilities" (abstract) \newline
	
	\underline{Findings}: (abstract)
	\newline Three steps of conducting accessibility assessment:
	\begin{enumerate}
		\item Preliminary review to quickly identify potential accessibility problems
		\item Conformance evaluation to determine whether a website meets established accessibility standards $\rightarrow$ Mainly focused on W3C Web Accessibility Initiative's (WAI) evaluation model.
		\item User testing to include real people with disabilities in a practical use
	\end{enumerate}
	
	\underline{	Future Work}: (abstract) \newline
	``Conclusions about actual barriers of web search engines and criteria of satisfaction for people with disabilities do not exist as of yet; the model is not tested so far."
	\\ \hline
	\cite{ sahib_accessible_2012} & & This paper also includes information on VI users and search engines since the subjects were required to collect online information on given tasks without specific guidelines or restrictions. \\ \hline
	\cite{ ivory_search_2004} & & \underline{PCA}: (p1) \newline
	\begin{enumerate}
		\item Formulating a query for an information need 
		\item Inspecting search results to identify relevant results
		\item Exploring potentially relevant pages to locate desired information
	\end{enumerate}
	* Users often revise queries and repeat these steps *
	\\ \hline
	\cite{ andronico_improving_2006} & & \underline{Prerequisites}: \newline
	``At the beginning of this research a preliminary study was performed concerning accessibility and usability of search tools and eight guidelines were formulated for designing search engine user interfaces" \newline
	
	\underline{Objective}:
	\newline - ``the derived guidelines were applied in modifying the source code of Google's interface while maintaining the same look and feel in order to demonstrate that with very little effort it is possible to make interaction easier more efficient and less frustrating for sightless individuals"
	\newline - ``the paper focuses on interface design and implementation."
	
	Includes results of survey on SE preference computer usage and query statistics (p567) \\ \hline
	
	\cite{ andronico_can_2004} & & This article surveys how conformed are the common search engines to WCAG 1.0 compliance requirements. \newline
	
	\underline{Results}: \newline
	``Of all tools analyzed only Google conformed to priority 1 of WCAG 1.0 [4] meaning that it satisfies a minimal level of accessibility (level A) whereas other search engines directories and meta-searches presented some priority 1 errors"
	\\ \hline
\end{longtable}

References mentioned in the \underline{Search Engines Appendix}:
\begin{center}
	\cite{ aliyu_google_2014}, \cite{ andronico_can_2004}, \cite{ andronico_improving_2006}, \cite{ baguma_web_2008}, \cite{ griol_voiceapp_2011}, \cite{ gurari_vizwiz_2018}, \cite{ hu_investigating_2015}, \cite{ ivory_search_2004}, \cite{leporini_designing_2004}, \cite{ lewandowski_accessibility_2012}, \cite{malik_oriental_2014}, \cite{ mele_beyond_2010}, \cite{ sahib_accessible_2012}, \cite{ sahib_evaluating_2015}, \cite{ small_your_2009}, \cite{ tsai_mobile_2010}, \cite{yang_improved_2012}
\end{center}

\newpage

\subsection{Navigation} \label{append_navigation}

Navigation, includes content for \textit{Physical World Navigation} (PWN), helping its users navigate around physical spaces. Citations involving PWN are included to show how different modalities can be achieved for applications other than web navigation. 

\begin{longtable}{|p{\refNumCell}|p{\nonPWNcellWidth}|p{\pwnCellWidth}|}  \hline
	\textbf{Ref \#} & \multicolumn{1}{c|}{\textbf{Cyber Navigation}} & \multicolumn{1}{c|}{\textbf{PWN}} \\ \hline
	
	\cite{imerciv_buzzclip_2019} & The Buzzclip, developed by iMerciv Inc., is a PWN device that allow users to detect overhead obstacles at a range of distances. The device is small enough to fit in the palm of the hand and can be attached to practically any piece of clothing. The device is also flexible enough to fit on canes and can be hand-held. \newline\newline The impressive thing about the Buzzclip is it's discreetness and silence since all information is transmitted via tactile vibrations. This device can be considered as a Tactile User Interface (TUI) for PWN. Additionally, the device offers upper body detection not commonly accounted for when using canes and guide dogs.& \multicolumn{1}{c|}{\checkmark} \\ \hline
	\cite{ismail_search_2010} \newline & Findings:
	\newline - users where disoriented as to where they were on the screen
	\newline - ``they did not know where the active window they were looking into" (p3)
	\newline - ``As stated by W3C the visually impaired became disoriented among windows due to the content spawning new windows without warning the user" (p3) & \\ \hline
	\cite{yang_improved_2012} \newline &  - Table 1 includes VI difficulties for Navigation
	\newline - SSEB may provide orientation navigation assistance and site maps to make users feel confident of where they are and what they are doing (p3)
	\newline - Andronico et al. further verified some of these guidelines and principles... added shortcuts to make navigation faster (p2)
	& \\ \hline
	\cite{ tsai_mobile_2010} &   - Baeza-Yates et. al. SE should use the concept of ``relevance feedback" which feedback the retrieval results for the first round query as context of relevance & \\ \hline
	\cite{ siekierska_internet-based_2008} & & \multicolumn{1}{c|}{\checkmark} \\ \hline
	\cite{ murphy_empirical_2008} & \underline{Issues with online navigation:}
	\newline - Navigating and filling in web-forms was reported as problematic task (p6) 
	\newline - Site timing out meant user lost their relative position (p6) 
	\newline - VI users want freedom on web and not to rely on sighted user (p6)
	\newline - JavaScript auto-refresh feature is problematic (p6)
	
	Navigation Tactics:
	\newline - ``Participants also tended to want more feedback concerning the spatial location of images within a web page" (p7)
	\newline - Gaining an overview of a web page can be a challenging process 
	\newline - VI use tab and arrow keys to manoeuvre the page

	Page Visualization: (p7)
	\newline - Screen Reader Users (SRU) visualize a vertical list of points and links 
	\newline - No spatial perception 
	\newline - This would guide their methodology for searching the page 
	\newline - ``A considerable demand was found to be placed upon short and long-term memory usage" (p7) 
	\newline - Expert users would `speed-read' through a page to get an overview of its contents
	& \\ \hline
	\cite{ hakobyan_mobile_2013} & - AudioBrowser [24] allows users to navigate webpages while on the move (p4). System was built on evidence found in [53]. & \\ \hline
	\cite{ mele_beyond_2010} & - WoW provides the whole information by conveying it in one single browsable page & \\ \hline
	\cite{ griol_voiceapp_2011} & \underline{Main objectives of the VoiceApp system:} (p2) 
	\newline - ``to adequately convey to users the logical structure and semantics of content in web documents"
	\newline - ``provide them with easy ways to select which parts of a document to listen to"
	& \\ \hline
	\cite{ trippas_spoken_2016} & \underline{Navigational Issue:}
	\newline - ``leaving users uncertain as to what they have covered the information space [22]" & \\ \hline
	\cite{ hu_investigating_2015} & \underline{Motivation:} \newline
	``To date no study has been conducted to examine how people with cognitive disabilities navigate in different content structures." (abstract)
	
	\underline{Objective:} \newline``an empirical study to investigate the impact of different search methods and content structures on the search behavior of people with cognitive disabilities" (abstract)
	& \\ \hline
	\cite{ baguma_web_2008} & - The requirements generated in this paper also highlight the need for user orientation while navigating pages & \\ \hline
	\cite{ sahib_evaluating_2015} &  - This paper also discusses TrailNote, an application that manages the search process for VI users & \\ \hline
	\cite{ sahib_investigating_2014} &  - Focus on Multi-session search tasks (abstract)
	
	- ``Multi-session tasks can be cognitively taxing for visually impaired users because the lack of persistence of screen readers causes the load on working memory to be high." (abstract)
	
	\underline{Results:}\newline
	- ``discuss the strategies observed among participants to resume the search" (abstract)
	
	Introduce the concept of a Search Trail:
	\newline - ``The search trail as shown in Figure 1 automatically records the queries that the user issues and the search results that are visited during a search session." (p5)
	\newline - ``users reported that they took notes during the search process." (p9)
	\newline - ``they would copy the URL of the page in a text le" (p9)
	
	& \\ \hline
	\cite{ sahib_accessible_2012} & \underline{Objective:} \newline 
	- ``This project examines how visually impaired people search for web-based information" (abstract)
	
	\underline{Method:} \newline
	- ``we investigated the search behaviour of 15 visually impaired and 15 sighted searchers while they completed complex search tasks online." (p1) \newline
	- ``Examples of complex searches include: planning travel to a previously unvisited country" \newline
	- ``gathering information on a medical condition" \newline
	- ``We studied 4 pairs of participants undertaking collaborative information seeking (CIS) tasks. Each pair comprised one sighted user and one visually impaired user who used a screen reader."
	& \\ \hline
	\cite{ ivory_search_2004} & \underline{Objective:} (p1)
	\newline - Examine the information consulted and time expended to make exploration decisions (PCA step 2)
	\newline - Examine the time expended or cost of exploring webpages (PCA step 3)
	
	\underline{Findings:}
	\newline - ``users leveraged page features to gauge the amount of effort that is required to explore search pages and made exploration decisions accordingly" (abstract)
	\newline - ``Users' desire to know additional page details varied based on their visual ability and the results' relevance" (abstract)
	& \\ \hline
	\cite{ dobrisek_voice-driven_2002} & \underline{Objective:}
	\newline - Homer Web Browser is a, ``small self-voicing Web browser designed for blind users is presented." (abstract)
	& \\ \hline
	\cite{ schmandt_audio_1998} & - To use the Audio Hallway application, the user travels up and down the Hallway by head motion passing ‘rooms’ alternately on the left and right sides.
	& \\ \hline
	\cite{ mack_inattentional_1998} & - This book discusses the lack of perception due to blindness when preceiving content online \newline
	
	Note: because VI people cannot explicitly see the object online and they are forced to create implicit abstractions of these elements.
	$\rightarrow$ as a result: the internet must be restructured so that they are able to generate explicit differences between important components online
	& \\ \hline
\end{longtable}


References mentioned in the \underline{Navigation Appendix}:
\begin{center}
	\cite{dobrisek_voice-driven_2002}, \cite{griol_voiceapp_2011}, \cite{hakobyan_mobile_2013}, \cite{ismail_search_2010}, \cite{ivory_search_2004}, \cite{mack_inattentional_1998}, \cite{mele_beyond_2010}, \cite{murphy_empirical_2008}, \cite{sahib_accessible_2012}, \cite{sahib_investigating_2014}, \cite{schmandt_audio_1998}, \cite{siekierska_internet-based_2008}, \cite{tsai_mobile_2010}, \cite{yang_improved_2012}
\end{center}

\newpage

\subsection{Usability / User Interface } \label{append_UI}

Usability / User Interface, includes content for \textit{Physical World Navigation} (PWN)

\begin{longtable}{|p{2.5cm}|p{7cm}|p{7cm}|c|}  \hline
	
	\textbf{Ref \#} & \multicolumn{1}{c|}{\textbf{Products}} & \multicolumn{1}{c|}{\textbf{Studies}} & \multicolumn{1}{c|}{\textbf{PWN}}\\ \hline
	
	\cite{frauenberger_mode_2005} & & \underline{Types of interfaces}: (p1)
	\newline - \textit{Visual}: Graphical User Interface
	\newline - \textit{Audio}: Auditory User Interface
	\newline - \textit{Touch}: Tactile User Interface \newline
	
	\underline{Interaction Principles of UI Designs}: (p2) \newline
	- \textit{Availability}: The required parts of the application need to be available at the right time and should imply correct usage. Mapping between intended user actions and user operations required.\newline 
	
	- \textit{Affordances}: Provide strong clues to the operation of things (e.g. knobs $\rightarrow$ turning, buttons $\rightarrow$ pushing). When used effectively the user knows what to do with no further instructions. \newline
	
	- \textit{Constraints}: Minimize the number of possible actions and give information about the correct usage of UI elements.\newline

	& \\ \hline
	
	\cite{frauenberger_mode_2005} continued.. & &
	
	continued..: (p2) \newline

	- \textit{Natural mapping}: If the relationship between controlling elements of an application and their results are natural to the user, it simplifies the learning process of the application and assists in recall. Natural mapping depends on physical analogies and cultural standards, and is therefore subjective to different user groups. \newline
	
	- \textit{Conceptual models}: By interacting with an application the user builds up a conceptual model of it. If this model is equivalent to the task model of the application it allows the user to predict the effects of their actions. \newline
	
	- \textit{Feedback}: Information about the result of their actions are sent back to the user and enables immediate control of the input.
	& \\ \hline

	\cite{menzi-cetin_evaluation_2017} & & \underline{Objective}:
	\newline - ``aims to evaluate the usability of a university website by visually impaired students" (abstract)
	
	Results from Survey: (p4)
	\newline - Majority use Internet Explorer as preferred web browser \newline
	
	\underline{Findings}:
	\newline - require library staff to get started (p5)
	\newline - find computer
	\newline - start computer
	\newline - digitizing printed material they wish to read
	& \\ \hline
	\cite{halimah_voice_2008} &  - Watch YouTube to learn something new
	\newline - Read Wikipedia for quick facts \newline
	
	``This paper highlights the Mg Sys Visi system that has the capability of access to World Wide Web by browsing in the Internet checking sending and receiving email searching in the Internet and listening to the content of the search only by giving a voice command to the system." & & \\ \hline
	\cite{yang_improved_2012} & & - Classification of difficulties for VI users: (Table 1)
	\newline - ``Leporini et al. proposed three guidelines for user interface design" (p2) \newline\newline
	[18 21] Although Google has a simple user interface and often is highly accessible it may be further improved to simplify interaction for visually impaired people when using screen readers
	& \\ \hline
	\cite{ tsai_mobile_2010} & & - On mobile devices Google displays more results with less description (p2) $\rightarrow$ good for quick overview and avoidance of repeating query entry \newline
	
	\underline{Results} : (p3)
	
	\textit{Create a two step query formulation:}
	\begin{enumerate}
		\item users to provide some keyword(s) as the queries
		\item users to select what kinds of retrieved documents they really want (ie. PDF website news images ...) (Step 2 is optional (p3))
	\end{enumerate}
	- Table 1 (p4) $\rightarrow$ specifies which information is important when asking a user to commit to a specific query specification (ie. Images: Size, Content Type, File Type, Colouration ...)
	& \\ \hline
	\cite{ aliyu_google_2014} & & Important quote - ``[13] pointed out that increase in cognitive activities required by a search tool reduces the chances of finding information" (p2) & \\ \hline
	\cite{ chen_detecting_2003} & & - \underline{Problem Summary}: Large computer-based webpages are too large to be viewed on small form factor devices like ``handheld computers personal digital assistants (PDAs) and smart phones". \newline
	
	\underline{Process - Two Step}:
	\newline - ``Page Analysis - analyzes the structure of a give web-page" 
	\newline - ``Page Splitting / Auto-positioning - splits the webpage into a two-level hierarchy" \newline
	
	Note: ``For a web-page not suitable for splitting an auto-positioning method (or scrolling-by-block) is used to provide a similar user experience."
	& \\ \hline
	\cite{ kapur_alterego:_2018} & \underline{AlterEgo}: HCI technology where the user must only mouth the intended words so that the electrical stimulus is sent to the muscles. This allows the user to interface with the computer without a sound. & & \\ \hline
	\cite{ small_your_2009} & & Explains the comparison between \textit{Net Naive} (new internet users) and \textit{Net Savvy} (internet users with prev experience). & \\ \hline
	\cite{ jayant_supporting_2011} & EasySnap, VI photographer application: \newline - ``an application that provides audio feedback to help blind people take pictures of objects and people and show that blind photographers take better photographs with this feedback." 
	\newline - Allows VI users to take, share (face-to-face or via social media), and browse pictures without the need to see. (p3) & & \\ \hline
	\cite{ wu_visually_2014} & & & \\ \hline
	\cite{ crossland_smartphone_2014} & & \underline{132 Total participants in survey:} 
	\newline - 81 total VI were as likely to use a smartphone or tablet as those with low VI 
	\newline - 59 of smartphone users found speech was helpful 
	\newline - 51 camera and screen as a magnier 
	\newline - 48 used an e-book reader & \\ \hline
	\cite{ siekierska_internet-based_2008} & \underline{Focus}: \newline
	``This chapter focuses on web-based tactile and audio-tactile maps for blind and visually impaired users developed within the Mapping for the Visually Impaired project"
	& & \multicolumn{1}{c|}{\checkmark} \\ \hline
	\cite{ macias_adaptability_2002} & \underline{Objective}:
	\newline - ``In this paper we present KAI (Accessibility Kit for the Internet) that considers both the user and the designer."
	\newline - `` KAI includes a mixed audio/touch browser (WebTouch) that enables selective reading of contents." \newline
	
	\underline{Secondary Objective}:
	\newline - ``KAI is based on a new language BML (Blind Markup Language) that helps authors to develop better structured pages." & & \\ \hline
	\cite{ macias_webtouch:_2004} & & \underline{Issues with online accessibility}:
	\newline - ``most of the browsers used to surf the net are thought to be managed by users without visual disabilities"\newline
	
	\underline{Objective}:\newline
	- ``Our research group has developed such a tool called KAI (Kit for the Accessibility to the Internet)"
	
	- In this paper we focus on WebTouch and its two modalities for surfing the net: voice and tactile skills. & \\ \hline
	\cite{ chiang_computer_2005} & & - GUIs are widely regarded as a major advance in human-computer interaction. 
	\newline - Their heavy dependence on visual cues for input and output presents a significant problem for visually disabled patients. (p3) $\rightarrow$ as a result the internet will only become more visual (which it has since 2005) \newline
	
	\underline{New developments on tools}: (p8)
	\newline - W3C: Web Content Accessibility Guidelines which have become an international standard
	for creating universally accessible Web-based products.
	\newline - Improvements for replacements of computer mouse
	\newline - a target mouse that provides auditory assistive feedback when the pointer enters or exits a target region
	& \\ \hline

	\cite{ murphy_empirical_2008} & & \underline{Reasons for not downloading files}:
	\newline - security
	\newline - file may contain virus 
	\newline - lack of training to download and interpret files 
	\newline - fear of system settings to be altered \newline
	
	\underline{Time before info overload/fatigue}: (p8)
	\newline - Less experienced users: 0.5 to 4 hrs 
	\newline - More experienced users: 5 hrs to entire day
	\newline - Partially sighted user: vary due to eye condition (ie. eyestrain) 
	\newline - ``voice of a screen reader could be overloading in itself" (p8) \newline
	
	Note: CAPTCHA programs could never be resolved by VI users (ie. Anti-bot software) \newline
	
	& \\ \hline

	\cite{ murphy_empirical_2008} continued... & & 
	continued...
	
	\underline{JAWS}:
	\newline - Not enough training from IT or VI users 
	\newline - Most users felt that training by a JAWS pro would be very beneficial (complex UI)
	\newline - JAWS according to users in the study is ``the best on the market" (p5)
	\newline - ``difficult to scan a page using JAWS" to gain overview of web page or locate an item of interest
	& \\ \hline
	
	\cite{ mele_beyond_2010} &  - ``spatial representation is processed by an amodal system" (abstract) $\rightarrow$ implying that when perceiving something we see it as whole (amodal perception) \newline
	
	WoW outputs its results ``in a top-down hierarchical sequence starting from the greatest ranking level website to the lowest in several results pages" & & \\ \hline
	\cite{ trippas_spoken_2016} & & \underline{Objective}: \newline
	``investigate how to present search results through a conversation over a speech-only communication channel where no screen is available" (abstract) \newline
	
	\underline{Advantages of speech-only system}: (p1)
	\newline - operating machinery [6 7]
	\newline - no screen or keyboard is available [24]
	\newline - users are on the move [16 21]
	\newline - using wearable devices [5]
	& \\ \hline
	\cite{ baguma_web_2008} & & \underline{Objective}: \newline
	``This paper presents Web design requirements that can improve the accessibility of such websites for PWDs [People with Disabilities] particularly the blind." (abstract) & \\ \hline
	\cite{ muwanguzi_coping_2012} & & \underline{Objective}: \newline
	``This study examined the usability challenges and emotional reactions blind college students experienced in accessing educational materials and communicating with professors and colleagues through online technologies." (abstract)	
	& \\ \hline
	\cite{ sahib_evaluating_2015} & \underline{Objective}: \newline
	``redesign the spelling-support mechanism using nonspeech sounds to address previously observed diculties in interacting with this feature." \newline
	
	\underline{Findings}:
	``the search interface was effective in supporting participants for complex information seeking and that the proposed interface features were accessible and usable with speech-based screen readers." (abstract)
	& & \\ \hline
	\cite{ andronico_improving_2006} & & This paper identifies the accessibility and usability issues (with references to other papers) (p123) & \\ \hline
	\cite{ dobrisek_voice-driven_2002} & This paper presents the Homer Web Browser (p1) $\rightarrow$ this application can be classified as an AUI \newline
	
	\underline{Five main modules}: \newline
	\textit{Input} $\rightarrow$ Voice, keyboard and mouse \newline
	\textit{Output} $\rightarrow$ Speech and non-speech sounds indicating location of pointer (mouse) (p3)
	& & \\ \hline
	\cite{ schmandt_audio_1998} & \underline{Objective}:  \newline
	``Audio Hallway a virtual acoustic environment for browsing collections of related audio files" (abstract) \newline
	
	Note: This paper presents the notion of creating a new (virtual) world for the user to interact with the search process and relevant information & & \\ \hline
\end{longtable}

References mentioned in the \underline{Usability Appendix}:
\begin{center}
	\cite{ aliyu_google_2014}, \cite{ andronico_improving_2006}, \cite{ baguma_web_2008}, \cite{ chen_detecting_2003}, \cite{ chiang_computer_2005}, \cite{ crossland_smartphone_2014}, \cite{ dobrisek_voice-driven_2002}, \cite{frauenberger_mode_2005}, \cite{halimah_voice_2008}, \cite{ jayant_supporting_2011}, \cite{ kapur_alterego:_2018}, \cite{ macias_adaptability_2002}, \cite{ macias_webtouch:_2004}, \cite{ mele_beyond_2010}, \cite{menzi-cetin_evaluation_2017}, \cite{ murphy_empirical_2008}, \cite{ muwanguzi_coping_2012}, \cite{ sahib_evaluating_2015}, \cite{ schmandt_audio_1998}, \cite{ siekierska_internet-based_2008}, \cite{ small_your_2009}, \cite{ trippas_spoken_2016}, \cite{ tsai_mobile_2010}, \cite{yang_improved_2012}
\end{center}

\newpage

\subsection{Information Accessibility} \label{append_info_access}

information accessibility, includes content for \textit{Physical World Navigation} (PWN)

\begin{longtable}{|c|p{7cm}|p{7cm}|c|}\hline
	
	\textbf{Ref \#} & \multicolumn{1}{c|}{\textbf{Products}} & \multicolumn{1}{c|}{\textbf{Studies}} & \multicolumn{1}{c|}{\textbf{PWN}}\\ \hline
	
	\cite{menzi-cetin_evaluation_2017} & & - VI users need to be informed of ``opportunities and events taking place on campus" (abstract)
	\newline - date related info was most difficult to acquire (generally displayed in a visual format; calendar) (abstract)\newline
	
	\underline{Difficulties encountered with school site:} (p5)
	\newline - Complexity of course registration
	\newline - Irregular reading order of the links at homepage
	\newline - Irregular listing of the announcements
	\newline - Failure to read visuals (posters etc.)
	\newline - Lack of direct access from web page to the target link
	&
	\\
	\hline
	\cite{ismail_search_2010} & & \underline{Objective}:
	\newline - to explore the challenges faced by the visually impaired learners in accessing virtual learning environment 
	\newline - to determine the suitable guidelines for developing a voice recognition browser that is accessible to the visually impaired 
	\newline - Developers are too esthetic oriented (p1) $\rightarrow$ ``Their main purpose is only to make their applications look fantastic and impressive"\newline
	
	\underline{Findings}:
	\newline - ``most of the existing web applications are not accessible" (p2)
	\newline - ``the user could not understand or grab the idea of information effectively" & \\ \hline
	\cite{halimah_voice_2008} & Includes translator that has the functionality to convert html codes to voice; voice to Braille and then to text again  \newline
	
	\underline{Results}:
	\newline - System can be used for other users of specially needs like the elderly and the physically impaired learners (not just VI).
	& & \\ \hline
	\cite{yang_improved_2012} &  - The objective of this paper was to design an accessible user interface for blind people the Specialized Search Engine for the Blind (SSEB) (p3)
	\newline - ``a minimum requirement is to ensure that everyone can understand the contents of any Webpage." (p1) & & \\ \hline
	\cite{ chen_detecting_2003} & \underline{Solution}: \newline Compartmentalize each Web page to a grid of thumbnails that can be individually accessed to reveal the data in detail.
	& & \\ \hline
	\cite{ kapur_alterego:_2018} & AlterEgo allows the user to request and receive information discreetly & & \\ \hline
	\cite{ siekierska_internet-based_2008} &  - ``production of maps for people with special needs poses new challenges"
	\newline - this article focuses on the access of physical world maps for people with VI
	& & \multicolumn{1}{c|}{\checkmark} \\ \hline
	\cite{ macias_adaptability_2002} & \underline{KAI includes 2 components}:
	\newline - BML
	\newline - WebTouch (more info on this in ``Macias - 2004 - WebTouch...") & & \\ \hline
	\cite{ macias_webtouch:_2004} & \underline{Issue with online accessibility}:
	\newline - ``the inaccessible design of the pages" \newline
	
	\underline{Components of KAI}: \newline
	$\rightarrow$ BML - a new markup language with accessibility features called Blind Markup Language
	
	$\rightarrow$ WebTouch - a multimodal browser taking blind people into special consideration
	& & \\ \hline
	\cite{ chiang_computer_2005} & & \underline{Accessibility can be achieved by}:
	\newline - translating the visual screen display into 
	\newline $\rightarrow$ auditory output (e.g. screen reading software with speech synthesizers) 
	\newline $\rightarrow$ tactile output (e.g. Braille display that echoes the screen display) 
	\newline $\rightarrow$ or a combination of the two modalities (p6)
	& \\ \hline
	\cite{ murphy_empirical_2008} & & - Generally no ALT Text (p7) 
	\newline - Some would copy and paste content to word processor to be analyzed later (p7) 
	\newline - Afraid of forgetting information! (p7)
	
	Note: The user is exposed to so much useless info that the are forced to listen and analyze carefully. (overload)
	
	- Less experienced blind users would send emails to friends and would visit websites recommended to them by friends (``blind-friendly sites") (p5) $\rightarrow$ Websites shouldn't be `blind-friendly' accessibility to information is something everyone should be able to do equally
	& \\ \hline
	\cite{ roentgen_impact_2009} & - This paper focuses on the use of electronic mobility devices in the physical world for obstacle detection orientation wayfinding and navigation systems \newline - BrailleNote GPS was found to be useful for the users to assist in learning a new space, therefore, Nonuse of the device (BrailleNote GPS) was noticed once the subject became familiar with the space
	& & \multicolumn{1}{c|}{\checkmark} \\ \hline
	\cite{ hakobyan_mobile_2013} & - This paper covers existing technologies that integrate the user to the physical world (ie. navigation, obstacle detection, space perception...). 
	
	- Also applications that aid navigation in the virtual world by compartmentalizing a webpage (AudioBrowser only).  & & \multicolumn{1}{c|}{\checkmark} \\ \hline
	\cite{ mele_beyond_2010} & & -``sonication method offers an effective tool able to transmit graphic information" (abstract) \newline - it is not enough to make a webpage accessible but also usable for those who do not interact with it naturally (p2) & \\ \hline
	
	\cite{ griol_voiceapp_2011} & VoiceApp enables to access and browse Internet by means of speech (abstract) \newline \newline
	\underline{Three components}: (Abstract) 
	\begin{enumerate}
		\item Voice Dictionary: allows the multimodal access to the Wikipedia encyclopedia
		\item Voice Pronunciations: developed to facilitate the learning of new languages by means of games with words and images
		\item Voice Browser: provides a fast and effective multimodal interface to the Google web search engine
	\end{enumerate}
	
	\underline{Conclusions}: (p9)
	\newline - creates a markup language to include relevant voice information from the webpage (VoiceXML)
	& & \\ \hline
	\cite{ trippas_spoken_2016} & & ``it is difficult to convey large amounts of information via audio without overloading the user's short-term memory [14 18 21]" & \\ \hline
	\cite{ yang_specialized_2007} & \textit{Reference} [4] in this paper, The Web Access Project, developed methods for adding captions and audio descriptions
	to movie clips
	& & \\ \hline
	\cite{ world_health_organization_who_world_2011} & & ``A study in 2008 found that five of the most popular social networking sites were not accessible to people with visual impairment (122)." (p185) & \\ \hline
	\cite{ hu_investigating_2015} & & Note: Though this article targets cognitive disorders the idea of preferring a format different than the standard is important to consider & \\ \hline
	\cite{ baguma_web_2008} & & \underline{Suggested Requirements}: (p8, p9, p10)
	\newline - a text only version of the website
	\newline - text alternatives for visual elements
	\newline - meaningful content structure in the source code
	\newline - skip navigation link(s) 
	\newline - orientation during navigation 
	\newline - to avoid the feeling of disorientation 
	\newline - ensure (tables frames and forms) are accessible
	\newline - test the website with keyboard only access
	\newline - use or convert documents into standard formats
	\newline - expand abbreviations and acronyms the first time they appear on a page. & \\ \hline
	\cite{ muwanguzi_coping_2012} & & \underline{Findings}:\newline
	``Schmetzke (2000) found that 23 out of 24 university websites audited in the United States did not comply with Web Accessibility Initiative guidelines (WAI 1999)."
	
	``Zaparyniuk and Montgomerie (2005) and Sloan Gregor Booth and Gibson (2002) conducted audits on web resources for 311 higher education institutions in Canada and in the U.K. respectively. They found that the educational resources 20 for academic staff and students contained design errors that seriously hindered accessibility and usability for individuals using adaptive software to access information." (p3) & \\ \hline
	\cite{ andronico_improving_2006} & & This paper identifies the accessibility and usability issues (with references to other papers) (p123) & \\ \hline
	
	\cite{aoda_2019} & An Act set out by the government of Ontario (Canada) detailing the standard required for ensuring an organization's product and services are accessible. The government of Ontario has set a goal to make the province fully accessible by 2025. & & \\ \hline
	
\end{longtable}

References mentioned in the \underline{Information Accessibility Appendix}:
\begin{center}
	\cite{aoda_2019}, \cite{ andronico_improving_2006}, \cite{ baguma_web_2008}, \cite{ chen_detecting_2003}, \cite{ chiang_computer_2005}, \cite{ griol_voiceapp_2011}, \cite{ hakobyan_mobile_2013}, \cite{halimah_voice_2008}, \cite{ hu_investigating_2015}, \cite{ismail_search_2010}, \cite{ kapur_alterego:_2018}, \cite{ macias_adaptability_2002}, \cite{ macias_webtouch:_2004}, \cite{ mele_beyond_2010}, \cite{menzi-cetin_evaluation_2017}, \cite{ murphy_empirical_2008}, \cite{ muwanguzi_coping_2012}, \cite{ roentgen_impact_2009}, \cite{ siekierska_internet-based_2008}, \cite{ trippas_spoken_2016}, \cite{ world_health_organization_who_world_2011}, \cite{ yang_specialized_2007}, \cite{yang_improved_2012}
\end{center}

\newpage

\subsection{Latency} \label{append_latency} 

Latency

\begin{longtable}{|c|p{7.5cm}|p{7.5cm}|}  \hline
	
	\textbf{Ref \#} & \multicolumn{1}{c|}{\textbf{Products}} & \multicolumn{1}{c|}{\textbf{Studies}} \\ \hline
	
	\cite{menzi-cetin_evaluation_2017} & & Table 6 captures latency by showing how much time each participant spent to complete each task \\ \hline
	\cite{ tsai_mobile_2010} &  \underline{Findings}: \newline
	``mobile users can spend less time to browse many irrelevant documents" (p1) & \\ \hline
	\cite{ kapur_alterego:_2018} & \underline{AlterEgo}: a non-invasive, non-vocal, HCI\newline
	
	Allows the user to input content to the computer at a faster rate & \\ \hline
	\cite{ jayant_supporting_2011} & \underline{EasySnap}, VI photographer app \newline 
	
	Major point: They are able to learn how to use the application very quickly [abstract] & \\ \hline
	\cite{ griol_voiceapp_2011} & Note: The importance of \underline{Wikipedia} as a source of quick information access & \\ \hline
	\cite{ sahib_investigating_2014} & This paper introduces a product called \textit{Search Trail} \newline
	
	- In order to reduce the time it takes to resume the session or a search, the user may revisit parts of the trail in order to resume (p12) & \\ \hline
	\cite{ ivory_search_2004} & & This paper is regarding speed of access
	
	Results:
	Show that blind users take MUCH longer to explore SE and webpages than sighted users (Table 2 p4)
	\\ \hline
	\cite{ andronico_improving_2006} & & This paper also considers the importance of a more efficient and quick user interface that reduces the latency in search engines \\ \hline
	\cite{ mack_inattentional_1998} & & This book addresses the lack of preception for VI users $\rightarrow$ in order for a VI user to create explicit perception of these online object they must spend significantly more time doing so \\ \hline
\end{longtable}

References mentioned in the \underline{Latency Appendix}:
\begin{center}
	\cite{ andronico_improving_2006}, \cite{ griol_voiceapp_2011}, \cite{ ivory_search_2004}, \cite{ jayant_supporting_2011}, \cite{ kapur_alterego:_2018}, \cite{ mack_inattentional_1998}, \cite{menzi-cetin_evaluation_2017}, \cite{ sahib_investigating_2014}, \cite{ tsai_mobile_2010}
\end{center}

\newpage

\subsection{Discreetness} \label{append_discreetness}

Discreetness
subcomponents (user voicing \& audio feedback)

\begin{longtable}{|c|p{16cm}|}  \hline
	
	\textbf{Ref \#} & \multicolumn{1}{c|}{\textbf{Products and Studies}} \\ \hline
	
	\cite{halimah_voice_2008} & The Mg Sys Vigi system is a web browser that accepts voice commands reducing its discreetness \\ \hline
	\cite{ismail_search_2010} & Note: This study focuses on Voice activated browsers which excludes discreetness \\ \hline
	\cite{ kapur_alterego:_2018} & This paper presents a wearable technology that allows the user to communicate with a comupter without actually speaking $\rightarrow$ the device picks up the signals from the brain to the vocal cords and forms the intended words on the computer\newline\newline Information can be relayed back via headphones thus ensuring a discreet experience \\ \hline
	\cite{humanware_braillenote} & The BrailleNote developed by Humanware\cite{humanware_braillenote}, is a braille display capable of: entering text through large buttons, and reading with a single line of refreshable braille displays. The unit houses USB and auxiliary sound as ports, as well as bluetooth capabilities with long battery life. The OS is available in different lanuages. \newline\newline The device is also available in several lengths that modify the length of the readable braille row. \\ \hline
	
\end{longtable}

References mentioned in the \underline{Discreetness Appendix}:
\begin{center}
	\cite{halimah_voice_2008}, \cite{humanware_braillenote}, \cite{ismail_search_2010}, \cite{ kapur_alterego:_2018}
\end{center}

\newpage

\subsection{Emotional Implications } \label{append_emotional_implications} 

Emotional Implications

\begin{longtable}{|c|p{7.5cm}|p{7.5cm}|}  \hline
	
	\textbf{Ref \#} & \multicolumn{1}{c|}{\textbf{Products}} & \multicolumn{1}{c|}{\textbf{Studies}} \\ \hline
	
	\cite{menzi-cetin_evaluation_2017} & & Intro: (p2)
	\newline - ``Usable products and contexts make people happy" (p1)
	\\ \hline
	\cite{ismail_search_2010} & \underline{Application}: \newline
	Voice Acitivated Browsers \newline
	
	\underline{Findings}:\newline
	Visually impaired learners feel disappointed (p3)
	& \\ \hline
	\cite{ tsai_mobile_2010} & & This paper claims that, there is a sensitive amount of time until the mobile user becomes frustrated (p2) \\ \hline
	\cite{ hersen_assertiveness_1995} & & Discusses the emotional implications and dependance of visually impaired people \\ \hline
	\cite{ small_your_2009} & & \underline{Findings}: \newline
	Discovers that increased brain activity of the Net Savvy users vs less activity in Net Naive while completing a google search $\rightarrow$ suggesting that computer usage is unnatural if the brain has to modify its activity
	\\ \hline
	\cite{ wu_who_2011} & This paper includes social media platforms such as Twitter & In the context of social media, this paper displays how influential specific groups of people are and what are the demographics of their follower base \\ \hline
	\cite{ jayant_supporting_2011} & & Sharing pictures and videos through social media has positive effects on people and their social circles \\ \hline
	\cite{ wu_visually_2014} & This paper includes data on technologies such as VoiceOver and iOS on the social media platform Facebook 
	& \underline{Findings}: \newline
	- Comparing the Facebook social networks of visually impaired users (VoiceOver Sample) vs. visually functioning users (iOS Sample) $\rightarrow$ Found strong similarities between the two in both network density size and usage 
	\newline - VoiceOver sample group found to receive more feedback from others online
	\\ \hline
	\cite{ crossland_smartphone_2014} & & The most frequently cited reason for NOT using these devices included:
	\begin{enumerate}
		\item Cost
		\item Lack of Interest
	\end{enumerate}
	\\ \hline
	
	\cite{ murphy_empirical_2008} & & - ``Half of the participants felt that they were missing out on a perceptual experience which they thought that the fully sighted community experience when viewing images" (p7)
	\newline - ``Navigating the Internet using a screen reader was reported to be a frustrating experience due to the lack of feedback received" (p5) \\ \hline
	\cite{ hakobyan_mobile_2013} & Mobile Assistive Technologies (MATs) & - Motivation for creating MATs is to help the user ``help individuals feel less stigmatized or labeled" (p2)
	\\ \hline
	\cite{ world_health_organization_who_world_2011} & & ``Online communities can be particularly empowering for those with hearing or visual impairments or autistic spectrum conditions (105) because they overcome barriers experienced in face-to-face contact." (p184)
	\\ \hline
	\cite{ muwanguzi_coping_2012} & & This paper also studies, ``the emotional reactions blind college students experienced in accessing educational materials and communicating with professors and colleagues through online technologies." (abstract) \\ \hline
	\cite{ sahib_investigating_2014} & This paper introduces a product called \textit{Search Trail} \newline
	
	The use of the \textit{trail} allows the user to know that their session is saved thus increasing confidence in the program & \\ \hline
	\cite{ andronico_improving_2006} & & This paper also address the need for less frustrating user intefaces that are user-oriented \\ \hline
\end{longtable}

References mentioned in the \underline{Emotional Implications Appendix}:
\begin{center}
	\cite{ andronico_improving_2006}, \cite{ crossland_smartphone_2014}, \cite{ hakobyan_mobile_2013}, \cite{ hersen_assertiveness_1995}, \cite{ismail_search_2010}, \cite{ jayant_supporting_2011}, \cite{menzi-cetin_evaluation_2017}, \cite{ murphy_empirical_2008}, \cite{ muwanguzi_coping_2012}, \cite{ sahib_investigating_2014}, \cite{ small_your_2009}, \cite{ tsai_mobile_2010}, \cite{ world_health_organization_who_world_2011}, \cite{ wu_visually_2014}, \cite{ wu_who_2011}
\end{center}

\newpage

\subsection{Visual Question Answering} \label{append_vqa} 

A sample of Visual Question Answering suveys

\begin{longtable}{|c|p{16cm}|}  \hline
	
	\textbf{Ref \#} & \multicolumn{1}{c|}{\textbf{Studies}} \\ \hline
	
	\cite{agrawal_vqa:_2017} & VQA
	\newline - might be useful for image to text descriptors of online visual media \\ \hline
	\cite{wang_explicit_2015} & VQA \\ \hline
	\cite{ gurari_vizwiz_2018} & VQA \\ \hline
\end{longtable}

References mentioned in the \underline{Visual Question Answering Appendix}:
\begin{center}
	\cite{agrawal_vqa:_2017}, \cite{ gurari_vizwiz_2018}, \cite{wang_explicit_2015}
\end{center}

\newpage

\subsection{Technologies \& Compliances } \label{append_tech_and_comp} 

Technologies \& Compliances

\begin{longtable}{|p{2.5cm}|p{7.5cm}|p{7.5cm}|}  \hline
	
	\textbf{Ref \#} & \multicolumn{1}{c|}{\textbf{Technology}} & \multicolumn{1}{c|}{\textbf{Compliance}} \\ \hline

	\cite{menzi-cetin_evaluation_2017} & \underline{Technologies}: 
	\newline - JAWS (with headset)
	\newline - Windows-Eyes
	& W3C \\ \hline
	\cite{ismail_search_2010} & \underline{Technologies}: 
	\newline - screen reader or screen magnifier
	\newline - JAWS $\rightarrow$ ``JAWS has limitation to describe images"
	& W3C \\ \hline
	\cite{halimah_voice_2008} & \underline{Mg Sys Visi}: specialized voice recognition browser
	\newline - originally designed and developed for the visually impaired learners
	
	System Composition (5 modules):
	\newline - Automatic Speech Recognition (ASR)
	\newline - Text-to-Speech (TTS)
	\newline - Search engine
	\newline - Print (Text-Braille)
	\newline - Translator (Text-to-Braille and Braille-to -Text)
	& \\ \hline
	\cite{yang_improved_2012} & \underline{Technologies}: 
	\newline - SSEB
	\newline - Screen Readers: JAWS or Big Eyes
	\newline - Personalized Search 
	& WCAG 1.0 (p2) $\rightarrow$ Verifiers considered that guidelines nos. 1, 4, 6, 12, and 14 were essential and should be given the highest priority \\ \hline
	\cite{ jacobson_cognitive_1998} &\underline{ Assistive devices for VI}: (Table 1 p2) 
	\newline - Long cane
	\newline - Hoople 
	\newline - Guide dog 
	\newline - Human assistant 
	\newline - Laser cane 
	\newline - Mowat sensor 
	\newline - Sonic Guide 
	\newline - Nottingham Obstacle Detector 
	\newline - NOMAD 
	\newline - Tactile displays/maps/arrays 
	\newline - Personal guidance system
	\newline - MoBIC 
	\newline - Atlas Strider 
	\newline - Talking signs 
	\newline - Auditory beacons 
	\newline - Electronic strips 
	\newline - Motion detectors 
	\newline - Pressure detectors 
	\newline - Bar code readers 
	\newline - Beacons 
	\newline - Braille/Auditory compass 
	\newline\newline ... continued on next page ... \newline
	& \\ \hline
	\cite{ jacobson_cognitive_1998} continued... &  
	\newline - Vision enhancing devices (monocular) 
	\newline - Infrared detectors 
	\underline{Physical Barriers in Navigation}: (Table 2 p3) 
	\newline - Pavement furniture 
	\newline - Cars parked on pavement (sidewalk) 
	\newline - Inability to read visual cues (e.g. street signs) 
	\newline - Construction/repair 
	\newline - Irregular uneven or broken surface 
	\newline - Crowds of people 
	\newline - Steps 
	\newline - Traffic lights without audible or pedestrian sequence 
	\newline - Weather 
	\newline - Lack of railings 
	\newline - Imperceptible kerb cuts (dropped kerbs) 
	\newline - Elevators 
	\newline - Distance 
	\newline - Door location 
	\newline - Door handles 
	\newline\newline ... continued on next page ... \newline
	& \\ \hline
	
	\cite{ jacobson_cognitive_1998}	continued... &  
	\newline - Nonstandard fixtures (shop front rails baskets and stalls) 
	\newline - Traffic hazards 
	\newline - Surface textures (lack of) 
	\newline - Overhead obstructions (overhanging signs cables vegetation) 
	\newline - Lack of cues (e.g. uniform open space) 
	\newline - Gradient 
	\newline\newline\underline{List Source}: Golledge and Stimpson (1997 p. 493).
	& \\ \hline
	
	\cite{ kapur_alterego:_2018} & AlterEgo & \\ \hline
	\cite{ wu_visually_2014} & \underline{Mentioned}:
	\newline - VoiceOver
	\newline - Facebook & \\ \hline
	\cite{ crossland_smartphone_2014} & \underline{Mentioned}:
	\newline - Smartphone
	\newline - Tablet
	\newline - Camera as a magnifier
	\newline - e-book reader & \\ \hline
	\cite{ chiang_computer_2005} & \underline{Technologies}: (p7)
	\newline - JAWS, by Freedom Scientific
	\newline - outSPOKEN, by ALVA Access Group
	\newline - HAL, by Dolphin Computer Access 
	\newline - IBM Home Page Reader, by IBM
	\newline - Narrator, by Microsoft Corp. \newline
	
	\underline{Common assistive technologies}: (p7)
	\newline - Screen magnifiers
	\newline - Screen readers
	\newline - Braille displays
	& W3C \\ \hline
	\cite{ murphy_empirical_2008} & \underline{Technologies}:
	\newline - JAWS (p5) 
	\newline - Windows Eyes (p5) 
	\newline - Supernova (p5) 
	\newline - ZoomText (p5) 
	\newline - Brailler (p7)
	\newline - Dictaphone (p7)
	\newline - Screen Readers
	& Web Content Accessibility Guidelines 1.0 (WCAG) $\rightarrow$ Studies (p2) show that these guidelines are insufficient. \\ \hline
	\cite{ roentgen_impact_2009} & \underline{Technologies}:
	\newline - BGPS - BrailleNote GPS
	\newline - UC - Ultracane
	\newline - TTTP - Teletact and Tom Pouce
	\newline - SP - Sonic Pathnder
	\newline - LC - Laser Cane
	\newline - No longer available Products:
	\newline - MS - Mowat Sensor
	\newline - SG - Sonicguide
	\newline - PS - Pathsounder
	& \\ \hline
	\cite{ hakobyan_mobile_2013} &  - Mobile Assistive Technologies (MATs) are devices used to sighted users as to escape the stigma of traditional assistive tool (ie. cane, screen-readers, walkers, guide dog...) (p2)
	\newline - Table 1 includes a list of assisted device, their intended usage, and their platform of operation. & \\ \hline
	\cite{ mele_beyond_2010} & WoW (WhatsOnWeb) is:
	\newline - ``an application tool based on new graph visualization algorithms" (p2)
	& \\ \hline
	\cite{ trippas_spoken_2016} & \underline{Technologies}:
	\newline - interactive information retrieval (IIR) (abstract)
	\newline - Spoken Conversational Search System (SCSS) (abstract)
	\newline - screen readers
	& \\ \hline
	\cite{ yang_specialized_2007} & & WCAG 1.0:
	\newline - contains 14 significant guidelines (p1 intro)
	\newline - intended for all web devs
	\\ \hline
	\cite{ world_health_organization_who_world_2011} & DEFINITION:
	``An \textit{assistive technology device} can be defined as ``any item, piece of equipment or product whether it is acquired commercially modified or customized, that is used to increase maintain or improve the functional capabilities of individuals with disabilities" (59)." (p105)
	& \\ \hline
	\cite{ baguma_web_2008} & \underline{Technologies}:
	\newline - Job Accessibility with Speech ( JAWS ) 8.0
	& \\ \hline
	\cite{ muwanguzi_coping_2012} & \underline{Technologies}:
	\newline - JAWS (p3)
	\newline - Braille displays (p3)
	\newline - Window-Eyes (p3)
	\newline - Magic (p3)
	\newline - Zoom Text (p3)
	& W3C (p3) \\ \hline
	\cite{ sahib_evaluating_2015} & “we particularly focus on implementing TrailNote” $\rightarrow$ a tool to support visually impaired searchers in managing the search process \newline
	
	\underline{Technologies}: (abstract)
	\newline - JAWS
	\newline - VoiceOver
	\newline - Window-Eyes
	\newline - Search Trail (as mentioned in previous papers)
	\newline - TrailNote & \\ \hline
	\cite{ lewandowski_accessibility_2012} & & W3C Web Accessibility Initiative's (WAI) \\ \hline
	\cite{ sahib_accessible_2012} & \underline{Mentioned}:
	\newline - Screen Readers & \\ \hline
	\cite{ andronico_improving_2006} & \underline{Technologies}: 
	\newline - JAWS (along with its theory of operation) (p8) $\rightarrow$ ``JAWS is a fairly complex program itself requiring considerable knowledge to be used with maximum proficiency" & \\ \hline
	\cite{ andronico_can_2004} & & WCAG 1.0 \\ \hline

	\cite{ harper_quest_2008} & & \underline{Objective}: \newline
	``This study raises awareness about issues of access in higher education" (abstract) \newline\newline
	Web Compliance: (p1)
	\begin{enumerate}
		\item Priority 1 (Single A) 
		\item Priority 2 (Double A) 
		\item Priority 3 (Triple A) $\rightarrow$ full compliance and accessibility
	\end{enumerate}
	
	- Accessibility Tips for Web Page features (found in Appendix A) \newline
	
	\underline{Results}: \newline
	One (1) university website (of 12) achieved \textit{Triple A Compliance} \newline
	$\rightarrow$ ``Studying this institution's implementation process we propose that other institutions might emulate this exemplary model to achieve greater website accessibility for all constituents" (abstract)\newline
	
	 ... continued on next page ... \newline
	\\ \hline
	
	\cite{ harper_quest_2008} continued... & & 
	Number of compliant sites: (p3) \newline
	None - 4 \newline
	Single A - 6 \newline
	Double A - 1 \newline
	Triple A - 1 \newline
	
	\underline{Mentioned}:
	\newline - WCAG (p1)
	\newline - ``The United States' Section 508 by law requires U.S. government websites to be accessible." (p1)
	\newline - The Wayback Machine, an Internet tool has the utility to note changes among
	archived versions of websites (p4)
	\\ \hline
\end{longtable}

References mentioned in the \underline{Technologies \& Compliances Appendix}:
\begin{center}
	\cite{ andronico_can_2004}, \cite{ andronico_improving_2006}, \cite{ baguma_web_2008}, \cite{ chiang_computer_2005}, \cite{ crossland_smartphone_2014}, \cite{ hakobyan_mobile_2013}, \cite{halimah_voice_2008}, \cite{ harper_quest_2008}, \cite{ismail_search_2010}, \cite{ jacobson_cognitive_1998}, \cite{ kapur_alterego:_2018}, \cite{ lewandowski_accessibility_2012}, \cite{ mele_beyond_2010}, \cite{menzi-cetin_evaluation_2017}, \cite{ murphy_empirical_2008}, \cite{ muwanguzi_coping_2012}, \cite{ roentgen_impact_2009}, \cite{ sahib_accessible_2012}, \cite{ sahib_evaluating_2015}, \cite{ trippas_spoken_2016}, \cite{ world_health_organization_who_world_2011}, \cite{ wu_visually_2014}, \cite{ yang_specialized_2007}, \cite{yang_improved_2012}
\end{center}
\end{landscape}

\bibliographystyle{ACM-Reference-Format}
\bibliography{./ms}

\end{document}